\documentclass[10pt]{article}

\usepackage{amsmath}
\usepackage{amssymb}
\usepackage{mathrsfs}
\usepackage{graphics}

\usepackage{duotenzor1p1}

\title{\textbf{The Operator Tensor Formulation of Quantum Theory}}
\author{Lucien Hardy\\
\textit{Perimeter Institute,}\\
\textit{31 Caroline Street North,}\\
\textit{Waterloo, Ontario N2L 2Y5, Canada}}

\begin{document}

\maketitle

\begin{abstract}
A typical quantum experiment has a bunch of apparatuses placed so that quantum systems can pass between them.  We regard each use of an apparatus, along with some given outcome on the apparatus (a certain detector click or a certain meter reading for example) as an \emph{operation}.  An operation (for example $\mathsf{B}_\mathsf{a_1}^\mathsf{b_2a_3}$) can have zero or more quantum systems inputted into it and zero or more quantum systems outputted from it. The operation $\mathsf{B}_\mathsf{a_1}^\mathsf{b_2a_3}$ has one system of type $\mathsf a$ inputted, and one system of type $\mathsf b$ and one of type $\mathsf a$ outputted.  We can wire together operations to form circuits. For example, $\mathsf{A}^\mathsf{a_1} \mathsf{B}_\mathsf{a_1}^\mathsf{b_2a_3} \mathsf{C}_\mathsf{b_2a_3}$. Each repeated integer label here denotes a wire connecting an output to an input of the same type.  As each operation in a circuit has an outcome associated with it a circuit represents a set of outcomes that can happen in a run of the experiment.  In the operator tensor formulation of quantum theory each operation corresponds to an \emph{operator tensor}.  For example the operation $\mathsf{B}_\mathsf{a_1}^\mathsf{b_2a_3}$ corresponds to the operator tensor $\hat{B}_\mathsf{a_1}^\mathsf{b_2a_3}$.  Further, the probability for a general circuit is given by replacing operations with corresponding operator tensors as in
\begin{equation}
\text{Prob}( \mathsf{A}^\mathsf{a_1} \mathsf{B}_\mathsf{a_1}^\mathsf{b_2a_3} \mathsf{C}_\mathsf{b_2a_3}) =
\hat{A}^\mathsf{a_1} \hat{B}_\mathsf{a_1}^\mathsf{b_2a_3} \hat{C}_\mathsf{b_2a_3}
\end{equation}
Repeated integer labels indicate that we multiply in the associated subspace, then take the partial trace over that subspace.  Operator tensors must be physical (namely, they must have positive input transpose and satisfy a certain normalization condition).
\end{abstract}

\tableofcontents

\section{Introduction}

One of Turing's great insights was the idea that a computational process can be broken up into a series of steps to be implemented by a machine.  The mathematical machinery that comes out of this insight constitutes a large part of the field of computer science.   Physics, on the other hand, has been largely dominated by the mathematics of differential equations as evidenced by Newtonian dynamics, Maxwell's equations, General Relativity, and the Schroedinger equation.  Recently, however, this has begun to change.  The techniques of computer science have seen  a dramatic and profound application to quantum theory in the guise of the new field of quantum information.   This has been accompanied by the importation of certain philosophical and methodological prejudices from computer science.  In particular, we can think of a quantum process as a computational process in which there is a user interface.  In fact, the idea of a user interface is very much in sympathy with the operational view of quantum theory where, on the users side of things, we make choices of measurements and take note of measurement outcomes.  This operational view of quantum theory is quite old.  The techniques of computer science provide a natural mathematical language for operationalism.  Operationalism is, here, a methodology. The success of such a methodology need not hinder an attempt to understand the world in ontological terms (where we ask what the underlying reality is).  Rather, there is every reason to hope the operational approach will ultimately bring us closer to such an ontological understanding.  It is likely that, one hundred years from now, physics will owe as much to the work of Turing as computer science does already.

In this paper we will present a new operational formulation of quantum theory - the operator tensor formulation - which provides an especially natural operational view of quantum theory.   To appreciate this new formulation we first need to consider how the old operational formulation of quantum theory works for circuits.

\subsection{Circuits and operations}

A circuit is formed by wiring together \emph{operations} such that there are no open wires left over.  For example,
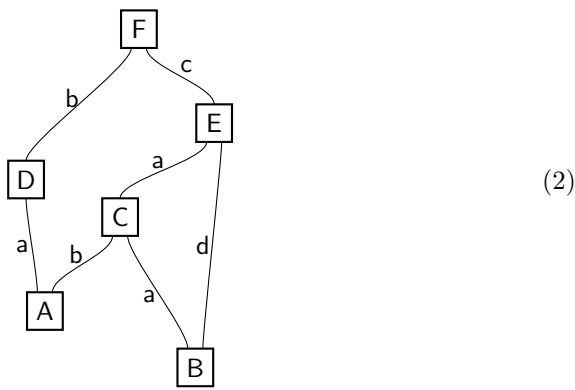
\begin{equation}\label{mediumcircuit}
\begin{Diagram}{0}{-1.8}
\Opbox[2]{A}{0,0}
\Opbox[2]{B}{8,-3}
\Opbox[2]{D}{-1,7}
\Opbox[2]{C}{4,5}
\Opbox[2]{E}{9,10}
\Opbox[2]{F}{5,15}
\wire{A}{D}{1}{1.5} \opsymbol{a} \wire{A}{C}{2}{1} \opsymbol[0,5]{b} \wire{B}{C}{1}{2} \opsymbol{a} \wire{B}{E}{2}{2}\opsymbol{d}
\wire{C}{E}{1.5}{1}\opsymbol{a} \wire{D}{F}{1.5}{1} \opsymbol{b} \wire{E}{F}{1.5}{2} \otherside\opsymbol[0,5]{c}
\end{Diagram}
\end{equation}
The operations are represented by boxes.  Systems pass along the wires.  The type of system (electron, atom, photon, \dots) is denoted by $\mathsf{a, b, c, \dots}$.   At each operation is a certain specified classical outcome (not shown in the figure) which is read off, for example, flashing lights, a meter, or something of this sort.  Hence, the circuit represents something that may happen - it could happen that we get the given outcome at each operation in the circuit.  We can give symbolic notation for this circuit
\begin{equation}
\mathsf{A^{a_1b_2}B^{a_3d_4}C_{b_2a_3}^{a_5} D_{a_1}^{b_6} E_{a_5d_4}^{c_7} F_{b_6c_7} }
\end{equation}
In this notation the wires are indicated by integers. For example, in the above expression $\mathsf{a_5}$ appears as a superscript on $\mathsf{C_{b_2a_3}^{a_5}}$ and a subscript on $\mathsf{E_{a_5d_4}^{c_7}}$ indicating a wire for a system of type $\mathsf a$ from an output of $\mathsf{C}$ to an input of $\mathsf{E}$.  The integers are merely labels and have no physical significance beyond this.  Since all the wiring information is carried by these integers, the order of the operations in the symbolic representation makes no difference.  We could, for example, equally represent the same circuit by
\begin{equation}
\mathsf{B^{a_3d_4}F_{b_6c_7}C_{b_2a_3}^{a_5}A^{a_1b_2} D_{a_1}^{b_6} E_{a_5d_4}^{c_7} }
\end{equation}
Relatedly, in the diagrammatic representation, the position (both vertical and horizontal) of the boxes on the page has no significance so long the wiring remains the same and the boxes maintain their orientation.

\subsection{Calculation in old formulation of quantum theory}\label{oldformulation}

How do we calculate the probability that all the specified outcomes at each of the boxes are seen?
The standard formulation of quantum theory, as usually applied to circuits of this sort, involves thinking in terms of a state evolving in time. Since we need to think of a state evolving in time first we must foliate the circuit \cite{hardy2009foliable}.
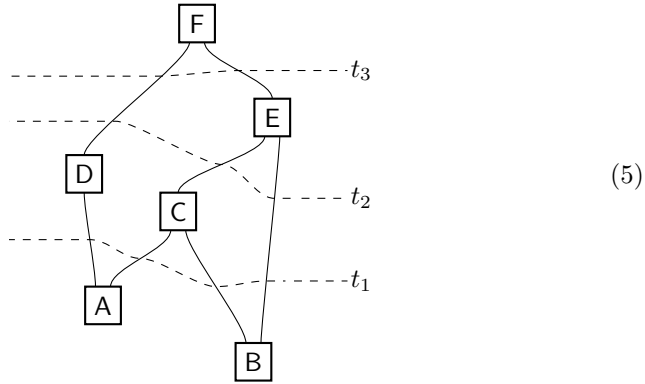
\begin{equation}
\begin{Diagram}{0}{-1.8}
\Opbox[2]{A}{0,0}
\Opbox[2]{B}{8,-3}
\Opbox[2]{D}{-1,7}
\Opbox[2]{C}{4,5}
\Opbox[2]{E}{9,10}
\Opbox[2]{F}{5,15}
\wire{A}{D}{1}{1.5} \wire{A}{C}{2}{1}  \wire{B}{C}{1}{2}  \wire{B}{E}{2}{2}
\wire{C}{E}{1.5}{1}\wire{D}{F}{1.5}{1}  \wire{E}{F}{1.5}{2}
\begin{foliation}{-5}{13}
\startfoliate{A}{D}{1}{1.5}\continuefoliate{A}{C}{2}{1}\continuefoliate{B}{C}{1}{2}\Finishfoliate{B}{E}{2}{2}{-0.2} \otherside\putlatex{\ensuremath{t_1}}
\Startfoliate{D}{F}{1.5}{1}{-0.2}\continuefoliate{C}{E}{1.5}{1}\Finishfoliate{B}{E}{2}{2}{0.2}
\otherside\putlatex{\ensuremath{t_2}}
\Startfoliate{D}{F}{1.5}{1}{0.2}\finishfoliate{E}{F}{1.5}{2}
\otherside\putlatex{\ensuremath{t_3}}
\end{foliation}
\end{Diagram}
\end{equation}
The foliation lines pick out the circuit analogue of space-time hypersurfaces.  In general, there is more than one such foliation for any given circuit. Now with respect to the chosen foliation we can write down an expression for the probability.
\begin{equation}\label{standardexpression}
\text{Prob}(\mathsf{A^{a_1b_2}B^{a_3d_4}C_{b_2a_3}^{a_5} D_{a_1}^{b_6} E_{a_5d_4}^{c_7} F_{b_6c_7} }) =
\text{Trace}\big(
\hat P_\mathsf{F}
[I_\mathsf{b}\otimes \$_\mathsf{E} ]
\circ
[\$_\mathsf{D}\otimes \$_\mathsf{C} \otimes I_\mathsf{d} ]
(\hat{\rho}_\mathsf{A}\otimes \hat{\rho}_\mathsf{B})\big)
\end{equation}
We get this expression by evolving the state prepared at time $t_1$ to time $t_2$, then to time $t_3$.   For the final step we take the trace with the operator associated with the final result $\mathsf F$.   Here $\hat{\rho}_\mathsf{A}$ is a positive operator (having trace less than or equal to one) associated with the \emph{preparation} $\mathsf A$.  Associated with the operation $\mathsf D$ is the trace non-increasing completely positive map $\$_\mathsf{D} $.  We will call an operation having both inputs and outputs, such as this, a \emph{transformation}.   By  $I_\mathsf{d}$ we denote the identity map on systems of type $\mathsf{d}$.  Finally, $\hat P_\mathsf{F}$ is a positive operator (that is less than or equal to the identity) associated with $\mathsf F$.  We will call operations, such as $\mathsf F$, having only input wires \emph{results}.   An important subtlety in this calculation is that we do not assume that the state is normalized after each time step.  The reason for this is that the normalization of the state is equal to the probability that the outcomes associated with the operations up to this time have happened and we do not want to lose this information.

There are three problems with doing calculation in the old formulation of quantum theory as evidenced by the calculation in (\ref{standardexpression}):
\begin{enumerate}
\item We need to choose an arbitrary foliation even though this is not part of the physics.
\item We have to pad the calculation with identity maps (such as $I_\mathsf{d}$) whenever two or more foliation lines intersect a given wire.
\item Our treatment is disunified in that we use different types of mathematical object for preparations (positive operator having trace less than or equal to one), transformations (trace non-increasing completely positive map), and results (positive operator less than or equal to the identity).
\end{enumerate}
We will resolve these problems in the operator tensor formulation.

\subsection{Calculation in operator tensor formulation}\label{intronewformalism}

In the operator tensor formulation there is an operator corresponding to each operation.  For example, the operator, $\hat C_\mathsf{b_2a_3}^\mathsf{a_5}$, corresponds to the operation, $\mathsf{ C}_\mathsf{b_2a_3}^\mathsf{a_5}$.  The operator $\hat C_\mathsf{b_2a_3}^\mathsf{a_5}$ is an element of the space of Hermitian operators acting on the Hilbert space $\mathcal{H}_\mathsf{b_2}\otimes\mathcal{H}_\mathsf{a_3}\otimes\mathcal{H}^\mathsf{a_5}$.  These operators have subscripts and superscripts associated with the inputs and outputs of the corresponding operation.  These subscripts and superscripts give the operators a tensor character so we will sometimes refer to them as \emph{operator tensors}.

In the operator tensor formulation the probability is given by the formula
\begin{equation}\label{operatortensorcalc}
\text{Prob}(\mathsf{A^{a_1b_2}B^{a_3d_4}C_{b_2a_3}^{a_5} D_{a_1}^{b_6} E_{a_5d_4}^{c_7} F_{b_6c_7} }) =
\hat A^\mathsf{a_1b_2}\hat B^\mathsf{a_3d_4}\hat C_\mathsf{b_2a_3}^\mathsf{a_5} \hat D_\mathsf{a_1}^\mathsf{b_6} \hat E_\mathsf{a_5d_4}^\mathsf{c_7} \hat F_\mathsf{b_7c_7}
\end{equation}
With standard tensor expressions we sum over repeated repeated labels.  However, the label in operator tensor corresponds to a Hilbert space rather than an integer and so we cannot simply sum over repeated labels here.  Rather, when we have a repeated label in an operator tensor expression we multiply in the appropriate subspace then take the partial trace in that subspace. We will explain in detail how this procedure works later on. Such repeated labels correspond to a wire such as $\mathsf{a_5}$.  Operators, such as $\hat C_\mathsf{b_2a_3}^\mathsf{a_5}$ for example, must be \emph{physical}.  We define an operator to be physical if it satisfies two properties:
\begin{description}
\item[\it It has positive input transpose:]  If we take the partial transpose over the part of the space associated with the input then we get a positive operator.  For example, $\hat C_\mathsf{b^T_2a^T_3}^\mathsf{a_5}\geq 0$.
\item[\it It has output trace less than the identity:] If we take the partial trace over the output we get an operator that is less than or equal to the identity. For example, $\hat C_\mathsf{b_2a_3}^\mathsf{a_5} \hat I_\mathsf{a_5} \leq I_\mathsf{b_2a_3}$.
\end{description}
These properties are imposed to guarantee that any probabilities we calculate are between 0 and 1.

Penrose showed how we can represent tensorial calculations diagrammatically.  This will be especially useful here.  We can give a diagrammatic version of (\ref{operatortensorcalc}) as follows
\begin{equation}\label{diagrammaticcalc}
\text{Prob}\left(
\begin{Diagram}{0}{-1.8}
\Opbox[2]{A}{0,0}
\Opbox[2]{B}{8,-3}
\Opbox[2]{D}{-1,7}
\Opbox[2]{C}{4,5}
\Opbox[2]{E}{9,10}
\Opbox[2]{F}{5,15}
\wire{A}{D}{1}{1.5} \opsymbol{a} \wire{A}{C}{2}{1} \opsymbol[0,5]{b} \wire{B}{C}{1}{2} \opsymbol{a} \wire{B}{E}{2}{2}\opsymbol{d}
\wire{C}{E}{1.5}{1}\opsymbol{a} \wire{D}{F}{1.5}{1} \opsymbol{b} \wire{E}{F}{1.5}{2} \otherside\opsymbol[0,5]{c}
\end{Diagram}
\right)
~~=~~
\begin{Diagram}{0}{-1.8}
\Dopbox[2]{A}{0,0}
\Dopbox[2]{B}{8,-3}
\Dopbox[2]{D}{-1,7}
\Dopbox[2]{C}{4,5}
\Dopbox[2]{E}{9,10}
\Dopbox[2]{F}{5,15}
\wire{A}{D}{1}{1.5} \opsymbol{a} \wire{A}{C}{2}{1} \opsymbol[0,5]{b} \wire{B}{C}{1}{2} \opsymbol{a} \wire{B}{E}{2}{2}\opsymbol{d}
\wire{C}{E}{1.5}{1}\opsymbol{a} \wire{D}{F}{1.5}{1} \opsymbol{b} \wire{E}{F}{1.5}{2} \otherside\opsymbol[0,5]{c}
\end{Diagram}
\end{equation}
The double boarder boxes on the right represent operators and the wires represent taking the partial trace over the corresponding Hilbert space (we will explain how to do this in detail later).  Hence, the figure on the right is an operator tensor calculation.

We note that, in the operator tensor formulation, for both (\ref{operatortensorcalc}) and (\ref{diagrammaticcalc}) the calculation (on the right hand side) looks the same as the circuit description (inside the argument on the left hand side) except that we have $\mathsf{A} \rightarrow \hat A$, $\mathsf{B}\rightarrow \hat B$, \dots and (in the diagrammatic version) we have double boarder boxes.  Further, with regard to the three points at the end of Sec.\ \ref{oldformulation}, we note that we do not need to impose a foliation, we do not need to pad the calculation with identities, and we give a unified treatment of preparations, transformations, and results (the operators corresponding to them must be physical).

\subsection{Previous work}

The operator tensor formulation was first presented in \cite{hardy2011reformulating} where it was used to aid a reconstruction of quantum theory from natural postulates. This paper is dedicated to presenting the operator tensor formulation (without the need to service a reconstruction effort) and consequently should be pedagogically clearer.  The closest thing in the literature to the operator tensor formulation of quantum theory is the \emph{quantum combs} approach of Chiribella, D'Ariano, and Perinotti \cite{chiribella2009theoretical}.  The relationship between the quantum combs approach and the present approach is discussed in detail in Sec.\ 8.5 of \cite{hardy2011reformulating} and, more briefly, below.

One of the main ideas in this paper is that we do not have to restrict ourselves to evolving a state forward in time in formulations of physical theories.  This idea has a long heritage.  In the context of quantum theory, Yakir Aharonov and co-workers have pursued such ideas for many years \cite{aharonov1964time}.   Most recently Aharonov, Popescu, Tollakson, and Vaidman have given the \lq\lq multi-time" formulation of quantum theory \cite{aharonov2009multiple} in which it is possible to have something analogous to a pure state pertaining to many times.   Relateldy, motivated by considerations from quantum gravity, Robert Oeckl has presented a general boundary formulation of quantum theory \cite{oeckl2003general}. In this it is possible to associate a quantum state with a general boundary of a space-time region.  This is also restricted to the pure state case.

Also motivated by quantum gravity, the present author set up the causaloid formalism \cite{hardy2005probability} (see also \cite{hardy2007towards,hardy2009quantum3,hardy2010bformalism}). This is a general framework for physical theories that does not demand that we have definite causal structure.  It was shown in \cite{hardy2005probability} how to formulate quantum theory in the causaloid framework.  This gave a formulation for the general (not just pure state) case in which we can associate mathematical objects that are analogous to quantum states with general space-time regions (actually this was a discrete situation so the space-time regions correspond to fragments of a circuit). Such mathematical objects can be combined by taking the \emph{causaloid product} to get the object for the union of the two space-time regions.

Subsequently, Chiribella, D'Ariano, and Perinotti gave the quantum combs formulation of quantum theory \cite{chiribella2009theoretical} which has much in common with the formulation presented in this paper (see also the related work of Gutoski and Watrous \cite{gutoski2006towards}).  In particular, in the quantum combs approach, an operator is associated with an operation.  This operator is equal to the input transpose of the operator introduced in Sec.\ \ref{intronewformalism} and is therefore positive.  In the quantum combs formalism it is possible to associate an operator with a fragment of a circuit and these operators can be combined using the \emph{link product}.   In the present framework we can combine operators in an analogous way to the way we combine tensors, by having repeated indices (which correspond to taking the partial trace over the corresponding space).  We call this the \emph{circuit trace}.  The circuit trace and, effectively, the link product are special cases of the causaloid product.   The main differences between the quantum combs formalism and the operator tensor formalism of this paper are (i)  we work with operators that are positive under input transpose, (ii) the operators here are given a tensorial structure which is exploited to simplify quite considerably the formulas, (iii) we are able to abandon the use of the tensor product symbol, $\otimes$, and do not have to pad equations with the identity, (iiii) the connection with duotensors (to be discussed later) makes it clear how we can go from a circuit to the corresponding operator tensor calculation for the probability simply by implementing $\mathsf{A_{a_1}^{b_2}}\rightarrow \hat A_\mathsf{a_1}^\mathsf{b_2}$, \dots.

Another reformulation of quantum theory that is clearly related to the present one is that of Leifer and Spekkens \cite{leifer2011formulating} (see also the earlier work of Leifer in \cite{leifer2006quantum}).  In this approach a conditional state, $\rho_{B|A}$, playing an analogous role to $P(B|A)$ in classical probability theory, is defined.  Quantum formulas can be written as instances of quantum belief propagation, $\rho_B= \text{Trace}_A( \rho_{B|A} \rho_A )$, which is analogous to classical belief propagation, $P(B) = \sum_R P(B|A) P(A)$.  The conditional state, $\rho_{B|A}$, is positive after taking the transpose over the space associated with $A$ (which is basically the same as the property of having positive input transpose considered in this paper).  Unlike the operator tensor formulation and the quantum combs formulation the approach of Leifer and Spekkens is not manifestly time asymmetric.  It would be interesting to understand better the relationship to the present approach, especially with respect to this issue of time asymmetry.

Prior to the work mentioned above, Markopoulou gave the quantum causal histories formulation \cite{markopoulou2000quantum}.  These causal graphs with matrix algebras at each vertex and completely positive maps on the edges.   Motivated by Markopoulou's work, Blute, Ivanov and Panangaden \cite{blute2003discrete} took the dual point of view, putting completely positive maps on the vertices and a similar point of view is adopted in this paper.    Markopoulou's work was very influential in getting people to think about putting quantum operations on a causal graph.

In this work a pictorial approach to understanding structure and doing calculations has been adopted.  This was motivated by the category theoretic approach of Abramsky and Coecke \cite{abramsky2004categorical}.   Pictures provide a powerful way of understanding the structure of calculations and their connection with the underlying physics.  The pictorialism revolution (see Coecke's \cite{coecke2010quantum}) initiated by Abramsky and Coecke is being more and more widely adopted.  In particular, see \cite{chiribella2010probabilistic} and the authors own papers \cite{hardy2010formalism}.  The use of pictures to represent calculations in this way goes back, of course, to Penrose's diagrammatic calculus for tensors \cite{penrose1971applications, penrose2005road}.  The diagrams in this paper were drawn using version 2 of the duotenzor package \cite{hardyduotenzor}.

Diagrammatic techniques have been used for representing \lq\lq tensor network states" \cite{vidal2006class} (see also \cite{verstraete2009matrix}), and recently Wood, Biamonte, and Cory have developed a tensor network graphical calculus for open quantum systems.

Sorkin and various co-workers have pursued the causal set approach to quantum gravity in which causal graphs are dynamically produced according to some growth dynamics \cite{sorkin1991spacetime}.  Recently, Johnston has shown how to put a discretized quantum field theory on a fixed causal set \cite{johnston2010quantum}.

At the heart of the operator tensor formulation of quantum theory are duotensors.  Duotensors are like tensors but with a bit more structure.  They were introduced by the author in \cite{hardy2010formalism} to provide way of formulating general probabilistic theories.  Such general probabilistic theories (sometimes described as the convex probabilities framework) have a long history going back to originally to Mackey and has been worked on by many others since including Ludwig \cite{ludwig1985axiomatic}, Davies and Lewis \cite{davies1970operational}, Gunson \cite{gunson1967algebraic}, Mielnik \cite{mielnik1969theory}, Araki \cite{araki1980characterization}, Gudder {\it et al.\ } \cite{gudder1999convex}, Foulis and Randall \cite{foulis1979empirical}, and Fivel \cite{fivel1994interference}.  More recent work on this framework includes \cite{hardy2001quantum, barrett2007information, barnum2011information, chiribella2010probabilistic, chiribella2010informational, hardy2009foliable, hardy2009operational, hardy2009operational2, hardy2010formalism}.

\section{The circuit model}

\subsection{Apparatuses and operations}

In quantum theory a typical experiment consists of a bunch of apparatuses, $\mathcal{A, B, C, \dots}$, placed such that apertures are matched up allowing systems (electrons, photons, atoms, \dots) to pass between them.  Typically we are interested in a \emph{use of an apparatus} during some time interval that systems are passing through (typically, this is determined by gating the use of the apparatus with respect to an external clock).  An apparatus may have settings (adjusted by knobs for example), and outcomes (read off meters, flashing lights, or the clicking of detectors for example).  We denote the outcome by $\mathsf{x}_\mathcal{A}$.

It is useful to associate what we will call an \emph{operation} with a use of an apparatus.  An operation has the following features:
\begin{enumerate}
\item identification of some apertures as \emph{inputs} and some apertures as \emph{outputs};
\item a given choice of \emph{setting}, $\mathsf{s(A)}$;
\item a given \emph{set of outcomes}, $\mathsf{o(A)}$.
\end{enumerate}
If the outcome that is seen is in the associated outcome set, i.e.\ $\mathsf{x}_\mathcal{A}\in \mathsf{o(A)}$, then we say that the operation \emph{happened} (by associating an outcome set with an operation rather than just a single outcome we, effectively, allow course-graining over outcomes).

The different types of system that may be inputted or outputted are represented by $\mathsf{a, b, c, \dots}$.  Sometimes we will use a single letter to denote a composite system (e.g.\ $\mathsf{d=ab}$).  Throughout this paper we will use both diagrammatic and symbolic notation.  We denote an operation by, for example,
\begin{equation}
\begin{Diagram}{0}{0}
\Opbox{A}{0,0}
\inwire[-5]{A}{1} \Opsymbol{a} \inwire{A}{2} \Opsymbol{b} \inwire[5]{A}{3}\Opsymbol{a}
\outwire[-4]{A}{1.5}\Opsymbol{c} \outwire[4]{A}{2.5} \Opsymbol{a}
\end{Diagram}
\text{diagramatic notation}
~~~~~
\mathsf{A_{a_1b_2a_3}^{c_4a_5}}
~~ \text{symbolic notation}
\end{equation}
In the diagrammatic notation the inputs are shown coming in from the bottom and the outputs are shown coming out from the top.  In the symbolic notation the inputs are shown as subscripts and the outputs are shown as superscripts.  In the symbolic notation we have integer subscripts on the type labels.  The purpose of these integers is to show where the wires go as will be discussed below.  Sometimes we will suppress the subscripts and superscripts and simply refer to the operation by, for example, $\mathsf A$.

For a given apparatus use with given identification of inputs and outputs and a given choice of settings we may still have different outcome sets and, therefore, different operations.  One interesting case is a set of such operations with disjoint outcome sets whose union is the set of all outcomes for this apparatus.  We will call this a \emph{complete set of operations}.  For this situation we will sometimes use the notation $\mathsf{A[i]}$ to represent the different operations in the complete set having associated (disjoint) outcome sets $\mathsf{o[i]}$ (whose union is the set of all outcomes).

\subsection{Wires and fragments}\label{wiresandfragments}

We can wire together operations to build \emph{fragments}. For example,
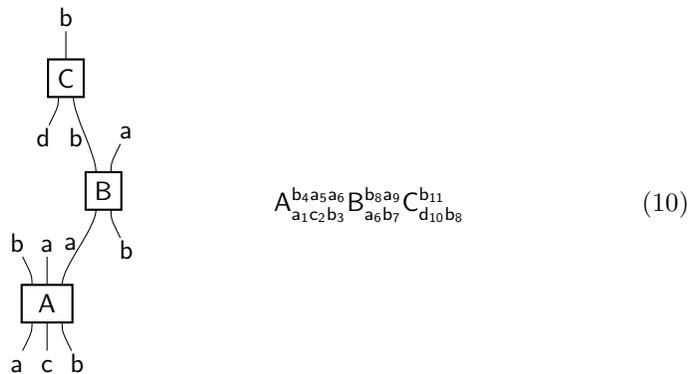
\begin{equation}\label{fragmentexample}
\begin{Diagram}{0}{-1.4}
\Opbox{A}{0,0} \Opbox[2]{B}{3,6} \Opbox[2]{C}{1,12}
\wire{A}{B}{3}{1} \opsymbol{a} \wire{B}{C}{1}{2} \opsymbol{b}
\inwire[-5]{A}{1} \Opsymbol{a}  \inwire{A}{2} \Opsymbol{c} \inwire[5]{A}{3} \Opsymbol{b}
\outwire[-5]{A}{1} \Opsymbol{b}  \outwire{A}{2} \Opsymbol{a}
\inwire[5]{B}{2} \Opsymbol{b}  \outwire[5]{B}{2} \Opsymbol{a}
\inwire[-5]{C}{1} \Opsymbol{d} \outwire{C}{1.5} \Opsymbol{b}
\end{Diagram}
~~~~~~~~~~~~~~~ \mathsf{ A_{a_1c_2b_3}^{b_4a_5a_6} B_{a_6b_7}^{b_8a_9} C_{d_{10}b_{8}}^{b_{11}} }
\end{equation}
We give the diagrammatic representation on the left and the symbolic representation on the right.  Note that, in the symbolic representation, repeated indices correspond to wires.  The outcome set associated with this fragment is $\mathsf{o(A)\times o(B) \times o(C)}$. If, in some rune of the experiment, the outcomes seen at the operations belong to the outcome sets at each operation comprising the fragment then we can say that the fragment happened for this run.   There are certain wiring rules when building fragments.
\begin{description}
\item[\it One wire:] At most one wire can be attached to any given input or output.
\item[\it Type matching:] We can only join outputs to inputs of the same type.
\item[\it No closed loops:] If we trace forward long wires from output to input going from one box to the next then we cannot get back to the box where we started.
\end{description}
Fragments are the most general types of experimental arrangement we consider.  Special cases include:
\begin{description}
\item[\it Circuits:] A circuit is a fragment having no open inputs or outputs.
\item[\it Preparations:] A preparation is a fragment having open outputs but no open inputs.
\item[\it Results:] A result is a fragment having open inputs but no open outputs.
\end{description}

We will sometimes represent fragments by a single letter, $\mathsf D$, for example.  We can write the example in (\ref{fragmentexample}) above as
\begin{equation}
\mathsf{E_{a_1c_2b_3b_7d_{10}}^{b_4a_5a_6a_9b_{11}} } ~~~ \{ \mathsf{ b_4, a_5 \rightarrow b_7, d_{10} ~~ a_9\rightarrow d_{10} } \}
\end{equation}
Here we have also specified the causal structure on the right.  The causal structure tells us which outputs can be fed (directly or indirectly) into which inputs.  For example, we can go from output $\mathsf{a_5}$ through some intermediate operations and into input $\mathsf{d_{10}}$.  Often we will simply refer to a fragment as, $\mathsf E$, for example without giving the inputs and outputs or causal structure information.

\subsection{Probabilities}

The most general object we can consider is a fragment.   In general a fragment will have open inputs and outputs.  Hence, we cannot expect the probability for the fragment to happen (i.e.\ that the outcomes on the apparatuses are in the outcome sets associated with the corresponding operations) to be independent of what is sent into the inputs (nor can we expect it to be independent of how we post-select on the outputs).  Hence we should be careful before attempting assigning a probability to a fragment - we may be unable to sensibly do so.  However, these considerations do not apply to circuits (this is the special case of a fragment when we have no open inputs or outputs).  Hence we make the following assumption:
\begin{quote}
{\bf Assumption 1.}  We can assign a probability with any given circuit (the probability that the circuit \lq\lq happens"), and this probability depends
only on the specification of the given circuit (the knob settings and outcome sets at the operations, and the wiring).
\end{quote}
In making this assumption we are assuming that we have appropriately characterized our apparatus uses as operations.  For example, if we had inadvertently omitted to include an input on the operation corresponding to some aperture on the apparatus then an adversary could send some system into this aperture and alter the probabilities.

An important consequence of this assumption is that probabilities factorize for a circuit composed of two disjoint circuits.  Thus, consider a circuit $\mathsf{AB}$ comprised of circuits $\mathsf A$ and $\mathsf B$.  We have
\begin{equation}\label{probfatorises}
\text{Prob}(\mathsf{AB})=\text{Prob}(\mathsf{A})\text{Prob}(\mathsf{ B})
\end{equation}
since
\begin{flalign*}
\text{Prob}&(\mathsf{AB}) = \text{Prob}\big(\mathsf{x_\mathcal{A}\in o(A), x_\mathcal{B}\in o(B)|\,sw(A), sw(B)}\big)  \\
 =& \text{Prob}\big(\mathsf{x_\mathcal{A}\in o(A)| x_\mathcal{B}\in o(B), sw(A), sw(B)}\big) \text{Prob}\big(\mathsf{x_\mathcal{B}\in o(B)|\, sw(A), sw(B)}\big)  \\
 =& \text{Prob}\big(\mathsf{x_\mathcal{A}\in o(A)|\, sw(A)}\big) \text{Prob}\big(\mathsf{ x_\mathcal{B}\in o(B)|\, sw(B)}\big)
\end{flalign*}
This property is essential for the framework we wish to set up.  A similar property was taken as a basic assumption by Chiribella, D'Ariano, and Perinotti in \cite{chiribella2010probabilistic}.

\section{Duotensors}

In this section we will set up the duotensor framework that provides the bridge between operations and operators.  To do this we need to define a notion of equivalence concerning operations.

\subsection{Equivalence}

To define a notion of equivalence first we define the $p(\cdot)$ function as follows
\begin{equation}
p(\alpha \mathsf{A} + \beta \mathsf{B} + \dots) := \alpha\text{Prob}(\mathsf{A}) + \beta\text{Prob}(\mathsf{B}) +\dots
\nonumber\end{equation}
for circuits $\mathsf A$, $\mathsf B$, \dots. and real numbers $\alpha$, $\beta$, \dots (these can be negative).  Note that the $p(\cdot)$ function is defined for linear sums of \emph{circuits}.  We cannot define it for general fragments since {\bf Assumption 1} only allows us to assign probabilities to circuits.

Before giving a general definition of equivalence first consider the following example.  We will say
\begin{equation}\label{equivalenceexampleone}
\mathsf{ \alpha A^{a_1} + \beta B^{a_1}  \equiv \gamma C^{a_1} +\delta D^{a_1} }
\end{equation}
if
\begin{equation}\label{equivalenceexampleonecondition}
p([\mathsf{ \alpha A^{a_1} + \beta B^{a_1}}] E_{a_1})  = \mathsf{ p([\gamma C^{a_1} +\delta D^{a_1}] E_{a_1}) } ~~~~\text{for all} ~~\mathsf{ E_{a_1}}
\end{equation}
In general we wish to consider expressions like the following
\begin{equation}
\text{expression} = \mathsf{ \alpha  + \beta A + \gamma B + \dots }
\end{equation}
where $\mathsf A$, $\mathsf B$, \dots are fragments.  Our definition of equivalence is the following.
\begin{quote}
{\bf Equivalence:}  We write
\begin{equation}
\text{expression}_1 \equiv \text{expression}_2
\end{equation}
if
\begin{equation}
p(\text{expression}_1 \mathsf{E}) \equiv p(\text{expression}_2 \mathsf{E})
\end{equation}
for any fragment $\mathsf E$ that makes the contents of the argument on both sides of this equation into a linear sum of circuits.
\end{quote}
Clearly (\ref{equivalenceexampleone}) is an example of this if (\ref{equivalenceexampleonecondition}) is satisfied.  Another example is the following.
\begin{equation}\label{equivalenceexampletwo}
\mathsf{ A} \equiv \text{Prob}(\mathsf{A}) ~~~ \text{for any circuit} ~~ \mathsf{A}
\end{equation}
To prove this note that
\begin{equation}
p(\mathsf{ A E} ) = p(\mathsf{A})p(\mathsf{E}) = p(\text{Prob}(\mathsf{A}) \mathsf{E})
\end{equation}
where $\mathsf E$ is an arbitrary circuit.  This example illustrates that equivalence is a weaker notion than equality.  A circuit is equivalent to a real number (its probability) even though these are two very different things (one exists in the laboratory, the other is a purely mathematical object).

In general there are two sorts of equivalence.
\begin{enumerate}
\item Each expression is a real number plus a linear combination of circuits:
\[
\alpha + \beta\mathsf{A} + \gamma \mathsf{B} + \dots \equiv \delta + \epsilon \mathsf{ C}+ \zeta \mathsf{D} + \dots
\]
where $\mathsf{A}$, $\mathsf B$, \dots, $\mathsf C$, $\mathsf D$, \dots, are all circuits.
\item Each expression is a linear combination of fragments
\[
\alpha\mathsf{A} + \beta \mathsf{B} + \dots \equiv \gamma\mathsf{ C}+ \delta \mathsf{D} + \dots
\]
where $\mathsf{A}$, $\mathsf B$, \dots, $\mathsf C$, $\mathsf D$, \dots, are all fragments having the same causal structure.
\end{enumerate}
An example of the first sort is given by (\ref{equivalenceexampletwo}) and an example of the second sort is given by (\ref{equivalenceexampleone}) (when (\ref{equivalenceexampleonecondition}) is satisfied).

\subsection{Fiducial preparations}

For any preparation it is possible to find a sum over a fiducial set of preparations to which the preparation is equivalent.  A fiducial set is essentially a basis.  We will notate the elements in a fiducial set of preparations for a system of type $\mathsf a$ in the following way
\begin{equation}
{}_{a_1}\!\mathsf{X}^\mathsf{a_1}
~~\Longleftrightarrow ~~
\begin{Diagram}{0}{-0.1}
\fidprep{X}{0,0} \inblack{X}{1} \Duosymbol{a} \outwire{X}{1} \Opsymbol{a}
\end{Diagram}
\end{equation}
where $a_1=1$ to $K_\mathsf{a}$ (this index labels the elements of the fiducial set).   On the right we give the diagramatic notation, on the right we give the equivalent symbolic notation.  We use sans serif font, $\mathsf a$, to denote the type.  We use normal math font, $a$, to denote the index running from $1$ to $K_\mathsf{a}$.  The reason for the odd placement of this index (as a pre-subscript) in the symbolic notation will become clear later.  In the diagrammatic notation we have a black dot on the fiducial elements.   We set things up such that we must match black dots with white dots when we do calculations.

Since we have a fiducial set of preparations, any preparation $\mathsf{A^{a_1}}$ can be written as
\begin{equation}
\mathsf{A^{a_1}} \equiv {}^{a_1}\!\!A \,\,\,{}_{a_1}\!\mathsf{X}^\mathsf{a_1}
~~~\Longleftrightarrow ~~~
\begin{Diagram}{0}{-0.2}
\Opbox[2]{A}{0,-7pt} \outwire{A}{1.5} \Opsymbol{a}
\end{Diagram}
~~ \equiv ~~
\begin{Diagram}{0}{-0.2}
\Duobox[2]{A}{0,-7pt} \fidprep{X}{5,0} \linkwb{A}{X}{1.5}{1}\duosymbol{a} \outwire{X}{1}\Opsymbol{a}
\end{Diagram}
\end{equation}
In the symbolic notation on the left we use Einstein summation convention over the repeated index $a_1$.  In the diagrammatic notation on the right we represent the summation over the index by the matching of a black and white dot.  Since we always match black and white dots it is convenient to define
\begin{equation}
\begin{Diagram}{0}{0}
\Duobox[2]{A}{0,-7pt} \fidprep{X}{5,0} \linkwb{A}{X}{1.5}{1}\duosymbol{a} \outwire{X}{1}\Opsymbol{a}
\end{Diagram}
~~ := ~~
\begin{Diagram}{0}{0}
\Duobox[2]{A}{0,-7pt} \fidprep{X}{4,0} \link{A}{X}{1.5}{1}\duosymbol{a} \outwire{X}{1}\Opsymbol{a}
\end{Diagram}
\end{equation}
Note that the vertical wires (running from outputs to inputs on operations) are physical - they correspond to systems passing between operations.  On the other hand, the horizontal links are mathematical - they correspond to summing over indices.  In the diagrams we write $a$ rather than $a_1$ because, even when we have many links (so we are summing over many indices) we can see where each link is by looking at the diagram without giving it an integer label.

The object ${}^{a_1}\!\!A$ is a list of numbers (it is an example of a duotensor) associated with the preparation.  It is the \emph{state} expressed with respect to the given fiducial set of preparations.  Diagrammatically the state, ${}^{a_1}\!\!A$, is represented by
\begin{equation}
\begin{Diagram}{0}{0}
\Duobox[2]{A}{0,0} \outwhite{A}{1.5} \Duosymbol{a}
\end{Diagram}
\end{equation}
In the case of quantum theory we will, later, see how we can also represent states by operators.

\subsection{Fiducial results}

Recall that a \emph{result} is a fragment having open inputs but no open outputs.  Just as with preparations,  we can set up a fiducial set of results which we can sum over (with some coefficients) to get an expression equivalent to any result.  We denote the fiducial results by
\begin{equation}
\mathsf{X}_\mathsf{a_1}^{a_1} ~~~\Longleftrightarrow ~~ ~~
\begin{Diagram}{0}{0.1}
\fideffect{X}{0,0} \outblack{X}{1}\Duosymbol{a} \inwire{X}{1}\Opsymbol{a}
\end{Diagram}
\end{equation}
where $a_1=1$ to $K_\mathsf{a}$.  Note that, once again, we put a black dot on the fiducial element. Then, for an arbitrary result for a system of type $\mathsf{a}$, we can write
\begin{equation}
\mathsf{B}_\mathsf{a_1}\equiv B_{a_1} \mathsf{X}_\mathsf{a_1}^{a_1}~~~~~\Longleftrightarrow  ~~~~~
\begin{Diagram}{0}{0}
\Opbox[2]{B}{0,0}\inwire{B}{1.5}\Opsymbol{a}
\end{Diagram}
~~\equiv~~
\begin{Diagram}{0}{0.07}
\fideffect{X}{0,-7pt}\Duobox[2]{B}{5,0}
\inwire{X}{1}\Opsymbol{a} \linkbw{X}{B}{1}{1.5} \duosymbol{a}
\end{Diagram}
\end{equation}
As before, we define
\begin{equation}
\begin{Diagram}{0}{0.07}
\fideffect{X}{0,-7pt}\Duobox[2]{B}{5,0}
\inwire{X}{1}\Opsymbol{a} \linkbw{X}{B}{1}{1.5} \duosymbol{a}
\end{Diagram}
~~ := ~~
\begin{Diagram}{0}{0.07}
\fideffect{X}{0,-7pt}\Duobox[2]{B}{4,0}
\inwire{X}{1}\Opsymbol{a} \link{X}{B}{1}{1.5} \duosymbol{a}
\end{Diagram}
\end{equation}

\subsection{Simple circuits}

We can form a circuit by following a preparation by a result for the same type of system.  We will first illustrate all of this using the diagrammatic notation (similar equations can always be written down in symbolic notation).
\begin{equation}\label{simplecircuitdiagram}
\begin{Diagram}{0}{-0.6}
\Opbox[2]{A}{0,0}\Opbox[2]{B}{0,4} \wire{A}{B}{1.5}{1.5}\opsymbol{a}
\end{Diagram}
~~\equiv~~
\begin{Diagram}{0}{-0.6}
\Duobox[2]{A}{0.5,-7.5pt} \fidprep{X2}{5,0} \linkwb{A}{X2}{1.5}{1}\duosymbol{a}
\fideffect{X}{5,4}\Duobox[2]{B}{9.5,4cm+7pt}
\linkbw{X}{B}{1}{1.5} \duosymbol{a}
\wire{X2}{X}{1}{1}\opsymbol{a}
\end{Diagram}
~~=~~
\begin{Diagram}{0}{-0.6}
\Duobox[2]{A}{0.5,-7.5pt} \fidprep{X2}{4,0} \link{A}{X2}{1.5}{1}\duosymbol{a}
\fideffect{X}{4,4}\Duobox[2]{B}{7.5,4cm+7pt}
\link{X}{B}{1}{1.5} \duosymbol{a}
\wire{X2}{X}{1}{1}\opsymbol{a}
\end{Diagram}
\end{equation}
Using the linearity of the $p(\cdot)$ function we have
\begin{equation}
\begin{Diagram}{0}{-0.6}
\Duobox[2]{A}{0.5,-7.5pt} \fidprep{X2}{5,0} \linkwb{A}{X2}{1.5}{1}\duosymbol{a}
\fideffect{X}{5,4}\Duobox[2]{B}{9.5,4cm+7pt}
\linkbw{X}{B}{1}{1.5} \duosymbol{a}
\wire{X2}{X}{1}{1}\opsymbol{a}
\end{Diagram}
~\equiv ~
\begin{Diagram}{0}{0}
\Duobox[2]{A}{0,0} \bbinsert[0.8]{g}{4,0} \duosymbol{a} \Duobox[2]{B}{8,0}
\link{A}{g}{1.5}{1}\link{g}{B}{1}{1.5}
\end{Diagram}
\end{equation}
where we define {\it the hopping metric}
\begin{equation}\label{hoppingdefinition}
\begin{Diagram}{0}{-0.1}
\bbmetric{g}{0,0}\duosymbol{a}
\end{Diagram}
~:=~
p\left(\!\!\!\!
\begin{Diagram}{0}{-0.6}
\fidprep{X2}{4,0}
\fideffect{X}{4,4}
\wire{X2}{X}{1}{1}\opsymbol{a}
\inblack{X2}{1} \Duosymbol{a}
\outblack{X}{1}\Duosymbol{a}
\end{Diagram}
\right)
~~~ \Leftrightarrow ~~~
\begin{Diagram}{0}{-0.6}
\fidprep{X2}{4,0}
\fideffect{X}{4,4}
\wire{X2}{X}{1}{1}\opsymbol{a}
\inblack{X2}{1} \Duosymbol{a}
\outblack{X}{1}\Duosymbol{a}
\end{Diagram}
~\equiv ~
\begin{Diagram}{0}{-0.1}
\bbmetric{g}{0,0}\duosymbol{a}
\end{Diagram}
\end{equation}
The equivalence on the right follows from the definition on the left by (\ref{equivalenceexampletwo}).
The hopping metric is a very important object in this work.  Its entries are the probabilities obtained when a fiducial preparation is followed by a fiducial result (and hence its entries are all positive).

\subsection{Black and white dots}

We must match black and white dots.  We can use the hopping metric to change a white dot into a black dot. Hence we have
\[
\begin{Diagram}{0}{-0.1}
\Duobox[2]{A}{0,0} \outblack{A}{1.5}
\end{Diagram}
\!\!\!\!=~
\begin{Diagram}{0}{-0.1}
\Duobox[2]{A}{0,0} \outwhite{A}{1.5} \bbmetric[0.7]{g}{3.7,0}
\end{Diagram}
~~~~~~~~~~
\begin{Diagram}{0}{-0.1}
\Duobox[2]{B}{0,0} \inblack{A}{1.5}
\end{Diagram}
~=
\begin{Diagram}{0}{-0.1}
\Duobox[2]{B}{0,0} \inwhite{A}{1.5} \bbmetric[0.7]{g}{-3.7,0}
\end{Diagram}
\]
Hence
\[
\begin{Diagram}{0}{-0.1}
\Duobox[2]{A}{0,0} \bbinsert[0.8]{g}{4,0} \duosymbol{a} \Duobox[2]{B}{8,0}
\link{A}{g}{1.5}{1}\link{g}{B}{1}{1.5}
\end{Diagram}
~=~
\begin{Diagram}{0}{-0.1}
\Duobox[2]{A}{0,0} \Duobox[2]{B}{5,0}
\linkwb{A}{B}{1.5}{1.5}\duosymbol{a}
\end{Diagram}
~=~
\begin{Diagram}{0}{-0.1}
\Duobox[2]{A}{0,0} \Duobox[2]{B}{5,0}
\linkbw{A}{B}{1.5}{1.5}\duosymbol{a}
\end{Diagram}
~:=~
\begin{Diagram}{0}{-0.1}
\Duobox[2]{A}{0,0} \Duobox[2]{B}{4,0}
\link{A}{B}{1.5}{1.5}\duosymbol{a}
\end{Diagram}
\]
We have
\[ \nwdots\bndots = \nndotslong = \nbdots\wndots \]
Hence, given the definitions we have set up, we can insert and delete pairs of black and white dots as we like.  Consistency requires
\begin{itemize}
\item \wwdots to be the inverse of \bbdots
\item \wbdots to be equal to the identity
\item \bwdots to be equal to the identity
\end{itemize}
The entries of the inverse, \wwdots, can be negative (in quantum theory they will be negative).  We can use the inverse to change a black dot into a white dot.  One example of this is for the fiducial elements themselves:
\begin{equation}
\begin{Diagram}{0}{0.1}
\fideffect{X}{0,0} \outblack{X}{1}\Duosymbol[65,0]{a} \inwire{X}{1}\Opsymbol{a}
\wwmetric[0.7]{g}{3.1,7pt}
\end{Diagram}
~=~
\begin{Diagram}{0}{0.1}
\fideffect{X}{0,0} \outwhite{X}{1}\Duosymbol{a} \inwire{X}{1}\Opsymbol{a}
\end{Diagram}
~~~~~~~~~~
\begin{Diagram}{0}{-0.1}
\fidprep{X}{0,0} \inblack{X}{1} \Duosymbol[-65,0]{a} \outwire{X}{1} \opsymbol{a}
\wwmetric[0.7]{g}{-3.1,-7pt}
\end{Diagram}
~=~
\begin{Diagram}{0}{-0.1}
\fidprep{X}{0,0} \inwhite{X}{1} \Duosymbol{a} \outwire{X}{1} \opsymbol{a}
\end{Diagram}
\end{equation}
These correspond to a sum over fiducial elements weighted by the appropriate elements in \wwdots.

\subsection{Symbolic notation for simple circuit}

In symbolic form the calculation in (\ref{simplecircuitdiagram}) can be written
\begin{equation}\label{probAB}
\mathsf{A^{a_1}B_{a_1}}  \equiv  {}_{a'_1}\!h^{a_1}\,\,{}^{a'_1}\!\!A B_{a_1} = A^{a_1}B_{a_1}= {}^{a_1}\!\!A \,\,{}_{a_1}\!B
\end{equation}
where we define
\begin{equation}
{}_{a'_1}h^{a_1} := p(\,{}_{a'_1}\!\mathsf{X}^\mathsf{a_1} \mathsf{X}_\mathsf{a_1}^{a_1})
\end{equation}
This is the hopping metric, \bbdots, in symbolic notation.  We define
\begin{equation}
A^{a_1} :=  {}_{a'_1}\!h^{a_1}\,\,{}^{a'_1}\!\!A  ~~~~~~~  {}_{a_1}\!B := {}_{a_1}\!h^{a'_1}\,\, B_{a'_1}
\end{equation}
so that the hopping metric can be used to \lq\lq hop" an index over.  It hops a subscript to the left (so it becomes a pre-subscript) and it hops a pre-superscript to the right (so it becomes a superscript).  The inverse of the hopping metric is ${}^{a'_1}\!h_{a_1}$ (this is the symbolic representation of \wwdots).

\subsection{Full decomposability}

Now we introduce our second assumption for the duotensor framework.
\begin{quote} {\bf Assumption 2: Full decomposability}. Any operation is equivalent to a linear combination of operations each of which consists of a result for each input and a preparation for each output.
\end{quote}
Each preparation and result in this decomposition can be expanded in terms of fiducial preparations and results respectively.  Hence, we do not lose any generality by choosing these results and preparations to belong to fiducial sets. Hence we can give a more useful equivalent formulation of the assumption.
\begin{quote}
{\bf Assumption 2: Full decomposability} Any operation $\mathsf{A_{a_1b_2\dots c_3}^{d_4e_5\dots f_6}}$ can be written, in diagrammatic notation, as
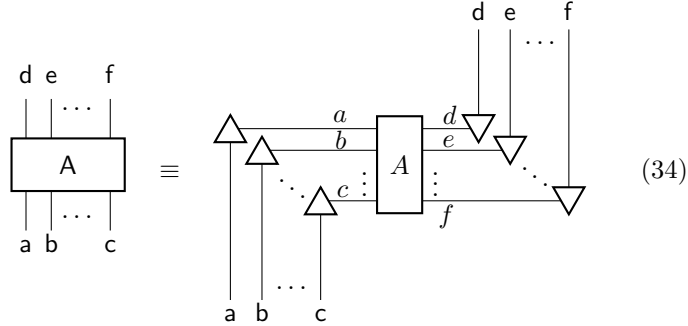
\begin{equation}\label{assumption2diagram}
\begin{Diagram}[1.4]{0}{0}
\Opbox[5]{A}{0,0}
\inwire{A}{1}\Opsymbol{a} \inwire{A}{2.2}\Opsymbol{b} \putlatex[30,15]{\ensuremath{\dots}} \inwire{A}{5}\Opsymbol{c}
\outwire{A}{1}\Opsymbol{d}\outwire{A}{2.2}\Opsymbol{e}\putlatex[30,-50]{\ensuremath{\dots}}   \outwire{A}{5}\Opsymbol{f}
\end{Diagram}
~~~\equiv ~~~
\begin{Diagram}[1.2]{0}{0}
\Duobox[5]{A}{0,0}
\putlatex[-44,-14]{\ensuremath{\vdots}} \putlatex[-135,-20]{\ensuremath{\ddots}}
\putlatex[45,-14]{\ensuremath{\vdots}} \putlatex[170,-5]{\ensuremath{\ddots}}
\linkedeffect[0.8]{A}{5}{c}{-3.5}{0}\duosymbol[-13,-4]{c} \thispoint{cbase}{-3.5,-6}\Opsymbol{c} \wire{cbase}{c}{1}{1}
\linkedeffect[0.8]{A}{2.2}{b}{-6.1}{0}\duosymbol[22,-3]{b}  \thispoint{bbase}{-6.1,-6}\Opsymbol{b} \wire{bbase}{b}{1}{1}
\linkedeffect[0.8]{A}{1}{a}{-7.5}{0}\duosymbol[41,-1]{a}  \thispoint{abase}{-7.5,-6}\Opsymbol{a} \wire{abase}{a}{1}{1}
\placelatex[-20, 16]{-4,-6}{\ensuremath{\dots}}
\linkedprep[0.8]{A}{1}{d}{3.5}{0}\duosymbol[4,1]{d}  \thispoint{dbase}{3.5,6}\Opsymbol[0,38]{d} \wire{d}{dbase}{1}{1}
\linkedprep[0.8]{A}{2.2}{e}{4.9}{0}\duosymbol[-17,-5]{e} \thispoint{ebase}{4.9,6}\Opsymbol[0,38]{e} \wire{e}{ebase}{1}{1}
\linkedprep[0.8]{A}{5}{f}{7.5}{0}\otherside\duosymbol[-57,-4]{f}  \thispoint{fbase}{7.5,6}\Opsymbol[0,38]{f} \wire{f}{fbase}{1}{1}
\placelatex[9, -16]{6,6}{\ensuremath{\dots}}
\end{Diagram}
\end{equation}
or, in symbolic notation,
\begin{equation}\label{keyassumption}
\mathsf{A_{a_1b_2\dots c_3}^{d_4e_5\dots f_6}} \,\,\equiv {}^{d_4e_5\dots f_6}\!A_{a_1b_2\dots c_3}\,\, \mathsf{X}_\mathsf{a_1}^{a_1} \mathsf{X}_\mathsf{b_2}^{b_2} \cdots \mathsf{X}_\mathsf{c_3}^{c_3} \,\,{}_{d_4}\!\mathsf{X}^\mathsf{d_4}{}_{e_5}\!\mathsf{X}^\mathsf{e_5}\cdots {}_{f_6}\!\mathsf{X}^\mathsf{f_6}
\end{equation}
\end{quote}
The assumption of full decomposability can be shown to be equivalent to the assumption of tomographic locality (also called local tomography) \cite{hardy2011reformulating} which states that it is possible to determine the state of a composite system by making measurements on the components. Another equivalent formulation is the assumption that $K_\mathsf{ab}=K_\mathsf{a}K_\mathsf{b}$.  Tomographic locality  is much studied \cite{araki1980characterization, bergia1980actual, wootters1986quantum, mermin1998quantum, hardy2010limited} and has been used as a postulate in many recent reconstructions of quantum theory from more natural postulates \cite{hardy2001quantum, dakic2009quantum, masanes2010derivation, chiribella2010informational}.  One final equivalent formulation of full decomposability (or tomographic locality) is \cite{hardy2011reformulating}
\begin{quote}
{\bf Assumption 2: Decomposability of wires}. A wire can be decomposed in the following way
\begin{equation}\label{identityfidsequation}
\begin{Diagram}{0}{0}
\thispoint{A}{-1,-3}\thispoint{B}{+1,+3}
\wire{A}{B}{1}{1}
\end{Diagram}
~\equiv~
\begin{Diagram}{0}{-0.1}
\fideffect[0.7]{X}{0,-0.15} \fidprep[0.7]{Y}{2,0.2}
\link{X}{Y}{1}{1}
\inwire{X}{1} \outwire{Y}{1}
\end{Diagram}
\end{equation}
\end{quote}
The equivalence of decomposability of wires to full decomposability of operations is proven in \cite{hardy2011reformulating}.  To gain some intuition into the reasons for this equivalence consider the fact that, if full decomposability is holds, then a wire can be regarded as an operation with one input and one output. By full decomposability it must be possible to expand it in terms of fiducial elements.  With a little thought it is clear that the duotensor associated with this operation is \wwdots and hence we can expand it as
\begin{equation}
\begin{Diagram}{0}{-0.1}
\fideffect[0.7]{X}{0,-0.15} \fidprep[0.7]{Y}{7,0.2}
\linkbwwb{X}{Y}{1}{1}
\inwire{X}{1} \outwire{Y}{1}
\end{Diagram}
\end{equation}
We can understand this as an alternative form of \wwdots is \wbdots which is equal to the identity which is the transformation we naturally associate with a wire.

\subsection{Duotensors}\label{duotensorssubsection}

Inserting black and white dots (with black next to the fiducial elements)
\begin{equation}\label{fulldecompexample}
\begin{Diagram}{0}{0}
\Opbox[2]{A}{0,0}
\inwire[-4]{A}{1}\inwire[4]{A}{2} \outwire[-4]{A}{1}\outwire[4]{A}{2}
\end{Diagram}
~~\equiv~~
\begin{Diagram}{0}{0}
\Duobox[2]{A}{0,0}
\fideffect[0.6]{1A}{-5.4,0.3}\inwire{1A}{1}
\fideffect[0.6]{2A}{-4,-0.6}\inwire{2A}{1}
\fidprep[0.6]{A1}{4,0.6}\outwire{A1}{1}
\fidprep[0.6]{A2}{5.4,-0.3}\outwire{A2}{1}
\linkbw{1A}{A}{1}{1} \linkbw{2A}{A}{1}{2}
\linkwb{A}{A1}{1}{1} \linkwb{A}{A2}{2}{1}
\end{Diagram}
\end{equation}
Therefore
\begin{equation}
\begin{Diagram}{0}{0}
\Duobox[2]{A}{0,0}\inwhite{A}{1}\inwhite{A}{2}\outwhite{A}{1}\outwhite{A}{2}
\end{Diagram}
\end{equation}
(with all white dots) provides the weights in the sum over fiducial elements.  This is an example of a \emph{duotensor}.  Duotensors are like tensors except that each index is associated with two bases (a fiducial basis for preparations and a fiducial basis for results). Duotensors transform like tensors but with respect to two bases (see \cite{hardy2009foliable, hardy2011reformulating} for a full discussion of how duotensors transform).  We can write duotensors symbolically, for example,
\begin{equation}
\begin{Diagram}{0}{0}
\Duobox[4]{A}{0,0}
\outwhite[-4]{A}{1.5}\Duosymbol{a} \outwhite{A}{2.5}\Duosymbol{b} \outblack[4]{A}{3.5}\Duosymbol{d}
\inblack[-6]{A}{1}\Duosymbol{b}    \inblack[-2]{A}{2}\Duosymbol{c}   \inwhite[2]{A}{3}\Duosymbol{b} \inwhite[6]{A}{4}\Duosymbol{c}
\end{Diagram}
~~\text{is written as}~~
  {}^{a_1b_2}_{b_3c_4} \! A^{d_5}_{b_6c_7}
\end{equation}
The following map is useful in converting between the diagramatic and symbolic notations.
\begin{equation}
{}_\bullet^\circ A^\bullet_\circ
\end{equation}
Importantly, duotensors are accompanied by a hopping metric for each system type which is used to convert between different forms of the duotensor.  In diagrammatic notation we can use \bbdots and \wwdots to change the colours of the dots.
\begin{equation}
\begin{Diagram}{0}{-0.1}
\Duobox[4]{A}{0,0}
\outwhite[-4]{A}{1.5}\Duosymbol{a} \outblack{A}{2.5}\Duosymbol{b} \outblack[4]{A}{3.5}\Duosymbol{d}
\inblack[-6]{A}{1}\Duosymbol{b}    \inwhite[-2]{A}{2}\Duosymbol{c}   \inblack[2]{A}{3}\Duosymbol{b} \inwhite[6]{A}{4}\Duosymbol{c}
\end{Diagram}
\!\! ~=~ \,
\begin{Diagram}{0}{-0.1}
\Duobox[4]{A}{0,0}
\outwhite[-4]{A}{1.5}\Duosymbol{a} \outwhite{A}{2.5}\Duosymbol[70,0]{b} \outblack[4]{A}{3.5}\Duosymbol{d}
\inblack[-6]{A}{1}\Duosymbol{b}    \inblack[-2]{A}{2}\Duosymbol[-70,0]{c}   \inwhite[2]{A}{3}\Duosymbol[-70,0]{b} \inwhite[6]{A}{4}\Duosymbol{c}
\wwmetric[0.8]{g1}{-3.8,0.53} \bbmetric[0.8]{g2}{-3.8,-0.53} \bbmetric[0.8]{g3}{3.8,0}
\end{Diagram}
\end{equation}
It follows from full decomposability that the duotensor with all black dots corresponds to the fiducial probabilities. For example,
\begin{equation}\label{allblackduotensor}
\begin{Diagram}{0}{0}
\Duobox[2]{A}{0,0}
\inblack{A}{1}\Duosymbol{a}\inblack{A}{2}\Duosymbol{b}
\outblack{A}{1}\Duosymbol{c}\outblack{A}{2}\Duosymbol{d}
\end{Diagram}
~=~
\text{Prob}\left(
\begin{Diagram}{0}{0}
\Opbox[2]{A}{0,0}
\fidprep[0.65]{1X}{-0.4,-3}\wire{1X}{A}{1}{1}\opsymbol{a}\inblack{1X}{1}\Duosymbol{a}
\fidprep[0.65]{2X}{0.4,-4.5}\wire{2X}{A}{1}{2}\otherside\opsymbol{b}\inblack{2X}{1}\Duosymbol{b}
\fideffect[0.65]{X1}{-0.4,4.5}\wire{A}{X1}{1}{1}\opsymbol{c}\outblack{X1}{1}\Duosymbol{c}
\fideffect[0.65]{X2}{0.4,3}\wire{A}{X2}{2}{1}\otherside\opsymbol{d}\outblack{X2}{1}\Duosymbol{d}
\end{Diagram}
\right)
\end{equation}
follows by inserting (\ref{fulldecompexample}) in the diagram on the right hand side, substituting in the hopping metric (\ref{hoppingdefinition}), and cancelling over matched black and white dots. Hence, we can obtain the duotensor with all black dots by process tomography using fiducial preparations and results.  We can then use the hopping metrics to convert to any other form.  In particular, the form with all white dots are the coefficients in the expansion of an operation in terms of fiducial elements.  This will be particularly important when we go over to operators where the duotensor with all white dots will be the weights in the decomposition of a operator tensor in terms of fiducial preparation operators and result operators (see Sec.\ \ref{Experimentaldeterminationsection}).   Another form of note is the \emph{standard form} in which we have white dots on the left and black dots on the right.  In this case, when we put the duotensors next to each other a black dot will always meet a white dot.  The duotensor associated with a preparation will have all black dots when in standard form and so constitutes a list of fiducial probabilities.

\subsection{General circuits}\label{generalcircuits}

Now we have set up the machinery sufficient to convert a general circuit into an equivalent duotensor calculation which, therefore, gives the probability for this circuit to happen.   To see this consider
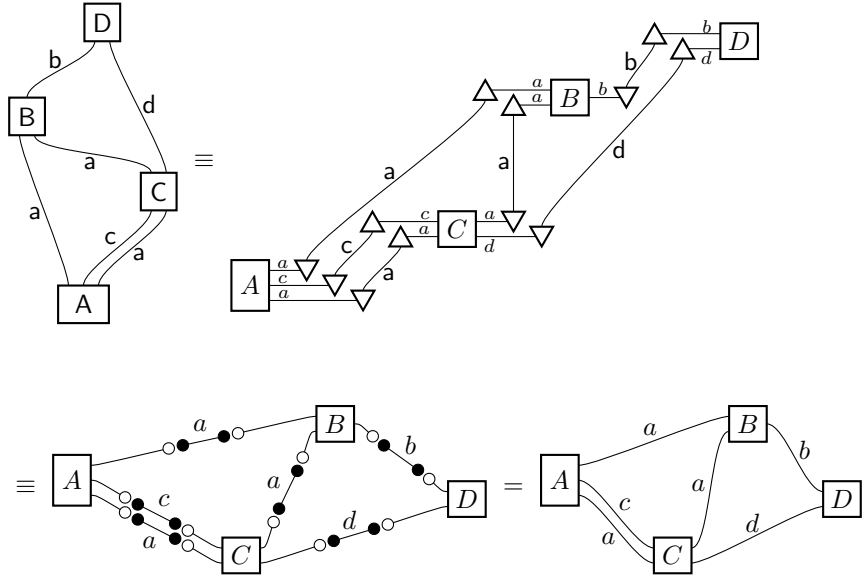
\begin{flalign*}
&\begin{Diagram}{0}{-2}
\Opbox{A}{0,0} \Opbox[2]{C}{4,6} \Opbox[2]{B}{-3,10} \Opbox[2]{D}{1,15}
\wire{A}{B}{1}{1}\opsymbol{a} \wire{A}{C}{2}{1}\opsymbol[0,6]{c} \wire{A}{C}{3}{2}\otherside\opsymbol{a} \wire{C}{B}{1}{2}\opsymbol{a}
\wire{C}{D}{2}{2}\otherside\opsymbol[4,0]{d} \wire{B}{D}{1.5}{1}\opsymbol[0,6]{b}
\end{Diagram}
~\equiv~
\begin{Diagram}{0}{-3}
\begin{move}{-7,5}
\Duobox{A}{0,0}
\linkedprep[0.7]{A}{1}{A1}{3}{0}\duosymbol[-3,-6]{\scriptstyle a}
\linkedprep[0.7]{A}{2}{A2}{4.5}{0}\duosymbol[-25,-6]{\scriptstyle c}
\linkedprep[0.7]{A}{3}{A3}{6}{0}\duosymbol[-44,-6]{\scriptstyle a}
\end{move}
\begin{move}{10,15}
\Duobox[2]{B}{0,0}
\linkedeffect[0.7]{B}{1}{1B}{-4.5}{0}\duosymbol[25,-6]{\scriptstyle a}
\linkedeffect[0.7]{B}{2}{2B}{-3}{0}\duosymbol[3,-6]{\scriptstyle a}
\linkedprep[0.7]{B}{1.5}{B15}{3}{0} \duosymbol[-3,-3]{\scriptstyle b}
\end{move}
\begin{move}{4,8}
\Duobox[2]{C}{0,0}
\linkedeffect[0.7]{C}{1}{1C}{-4.5}{0}\duosymbol[25,-6]{\scriptstyle c}
\linkedeffect[0.7]{C}{2}{2C}{-3}{0}\duosymbol[3,-6]{\scriptstyle a}
\linkedprep[0.7]{C}{1}{C1}{3}{0} \duosymbol[-3,-6]{\scriptstyle a}
\linkedprep[0.7]{C}{2}{C2}{4.5}{0} \otherside\duosymbol[-25,3]{\scriptstyle d}
\end{move}
\begin{move}{19,18}
\Duobox[2]{D}{0,0}
\linkedeffect[0.7]{D}{1}{1D}{-4.5}{0}\duosymbol[25,-6]{\scriptstyle b}
\linkedeffect[0.7]{D}{2}{2D}{-3}{0}\otherside\duosymbol[3,3]{\scriptstyle d}
\end{move}
\wire{A1}{1B}{1}{1}\opsymbol{a} \wire[0.4]{A2}{1C}{1}{1}\opsymbol{c} \wire[0.4]{A3}{2C}{1}{1}\otherside\opsymbol{a}
\wire{C1}{2B}{1}{1}\opsymbol{a} \wire{C2}{2D}{1}{1}\otherside\opsymbol{d} \wire[0.4]{B15}{1D}{1}{1}\opsymbol[-3,0]{b}
\end{Diagram}
\\
&\equiv~
\begin{Diagram}[1]{0}{0}
\Duobox{A}{-3,0} \Duobox[2]{B}{11,3} \Duobox[2]{C}{6,-4}  \Duobox[2]{D}{18,-1}
\linkwbbw[0.6]{A}{B}{1}{1}\duosymbol{a} \linkwbbw[0.6]{A}{C}{2}{1}\duosymbol{c} \linkwbbw[0.6]{A}{C}{3}{2}\otherside\duosymbol{a}
\linkwbbw[0.6]{C}{B}{1}{2}\duosymbol{a} \linkwbbw[0.6]{C}{D}{2}{2}\duosymbol{d} \linkwbbw[0.6]{B}{D}{1.5}{1}\duosymbol{b}
\end{Diagram}
~=~
\begin{Diagram}{0}{0}
\Duobox{A}{0,0} \Duobox[2]{B}{10,3} \Duobox[2]{C}{6,-4}  \Duobox[2]{D}{15,-1}
\link{A}{B}{1}{1}\duosymbol{a} \link{A}{C}{2}{1}\duosymbol{c} \link{A}{C}{3}{2}\otherside\duosymbol{a}
\link{C}{B}{1}{2}\duosymbol{a} \link{C}{D}{2}{2}\duosymbol{d} \link{B}{D}{1.5}{1}\duosymbol{b}
\end{Diagram}
\end{flalign*}
In the first step we have replaced each operation with its fully decomposed form using (\ref{assumption2diagram}).
At this stage we have a linear sum of products of fiducial circuits.  In the second step we apply the $p(\cdot)$ function and use the definition of the hopping metric in (\ref{hoppingdefinition}).  We also make use of the property that the $p(\cdot)$ function factorizes over disjoint circuits (since probabilities factorize over disjoint circuits as in (\ref{probfatorises})).

The bottom line is
\begin{equation}
\text{Prob}\left(
\begin{Diagram}{0}{-2}
\Opbox{A}{0,0} \Opbox[2]{C}{4,6} \Opbox[2]{B}{-3,10} \Opbox[2]{D}{1,15}
\wire{A}{B}{1}{1}\opsymbol{a} \wire{A}{C}{2}{1}\opsymbol[0,6]{c} \wire{A}{C}{3}{2}\otherside\opsymbol{a} \wire{C}{B}{1}{2}\opsymbol{a}
\wire{C}{D}{2}{2}\otherside\opsymbol[4,0]{d} \wire{B}{D}{1.5}{1}\opsymbol[0,6]{b}
\end{Diagram}
\right)
~ = ~
\begin{Diagram}{0}{0}
\Duobox{A}{0,0} \Duobox[2]{B}{10,3} \Duobox[2]{C}{6,-4}  \Duobox[2]{D}{15,-1}
\link{A}{B}{1}{1}\duosymbol{a} \link{A}{C}{2}{1}\duosymbol{c} \link{A}{C}{3}{2}\otherside\duosymbol{a}
\link{C}{B}{1}{2}\duosymbol{a} \link{C}{D}{2}{2}\duosymbol{d} \link{B}{D}{1.5}{1}\duosymbol{b}
\end{Diagram}
\end{equation}
This clearly follows for any circuit.  In other words the probability for a circuit is given by a duotensor diagram that looks the same (except that the font has changed and it has been rotated through $90^\circ$).

\section{The machinery of operator tensors}

\subsection{Operator tensors}

We will now set up another mathematical structure, namely that of operator tensors.  In the next section we will see how to use these to reformulate quantum theory.   We define complex Hilbert spaces ${\cal H}_\mathsf{a_1}$, ${\cal H}_\mathsf{b_2}$, \dots having dimensions $N_\mathsf{a}$, $N_\mathsf{b}$, \dots. We define complex Hilbert spaces ${\cal H}^\mathsf{a_1}$, ${\cal H}^\mathsf{b_2}$, \dots having dimensions $N_\mathsf{a}$, $N_\mathsf{b}$, \dots. We define
\begin{equation}
{\cal H}_\mathsf{a_1b_2\dots c_3}^\mathsf{d_4e_5\dots f_6} := {\cal H}_\mathsf{a_1} \otimes {\cal H}_\mathsf{b_2} \otimes \dots \otimes {\cal H}_\mathsf{c_3}
\otimes {\cal H}^\mathsf{d_4} \otimes {\cal H}^\mathsf{e_5} \otimes \dots \otimes {\cal H}^\mathsf{f_6}
\end{equation}
We define
\[ {\cal V}_\mathsf{a_1b_2\dots c_3}^\mathsf{d_4e_5\dots f_6}   \]
as the space of \emph{Hermitian} operators acting on ${\cal H}_\mathsf{a_1b_2\dots c_3}^\mathsf{d_4e_5\dots f_6}$.
We write
\begin{equation}
\hat A_\mathsf{a_1b_2\dots c_3}^\mathsf{d_4e_5\dots f_6} ~~~~~\Leftrightarrow~~~~~
\begin{Diagram}[1.4]{0}{0}
\Dopbox[5]{A}{0,0}
\inwire{A}{1}\Opsymbol{a} \inwire{A}{2.2}\Opsymbol{b} \putlatex[30,15]{\ensuremath{\dots}} \inwire{A}{5}\Opsymbol{c}
\outwire{A}{1}\Opsymbol{d}\outwire{A}{2.2}\Opsymbol{e}\putlatex[30,-50]{\ensuremath{\dots}}   \outwire{A}{5}\Opsymbol{f}
\end{Diagram}
\end{equation}
for elements of ${\cal V}_\mathsf{a_1b_2\dots c_3}^\mathsf{d_4e_5\dots f_6}$. We call operators having only superscripts \emph{preparation operators} and we call operators having only subscripts \emph{result operators}.

\subsection{Fiducial operators}

We introduce a fiducial (spanning) set of preparation operators for ${\cal V}^\mathsf{a}$
\begin{equation}
{}_{a_1}\!\hat{X}^\mathsf{a_1}~~\Longleftrightarrow ~~
\begin{Diagram}{0}{-0.1}
\dfidprep{X}{0,0} \inblack{X}{1} \Duosymbol{a} \outwire{X}{1} \opsymbol{a}
\end{Diagram}
\end{equation}
where $a_1=1$ to $K_\mathsf{a}$.   We have that $K_\mathsf{a}=N_\mathsf{a}^2$ as this is the number of linearly independent operators spanning the space of Hermitian operators on an $N_\mathsf{a}$ dimensional complex Hilbert space.  Similarly, we introduce a fiducial (spanning) set of result operators for the space ${\cal V}_\mathsf{a_1}$
\begin{equation}
\hat{X}_\mathsf{a_1}^{a_1}
~~\Longleftrightarrow ~~
\begin{Diagram}{0}{0.1}
\dfideffect{X}{0,0}  \outblack{X}{1}\Duosymbol{a} \inwire{X}{1}\Opsymbol{a}
\end{Diagram}
\end{equation}
where $a_1=1$ to $K_\mathsf{a}$.

\subsection{Full decomposability of operator tensors}

Since the Hilbert spaces we are using are complex we have
\begin{equation}
{\cal V}_\mathsf{a_1b_2\dots c_3}^\mathsf{d_4e_5\dots f_6} := {\cal V}_\mathsf{a_1} \otimes {\cal V}_\mathsf{b_2} \otimes \dots \otimes {\cal V}_\mathsf{c_3}
\otimes {\cal V}^\mathsf{d_4} \otimes {\cal V}^\mathsf{e_5} \otimes \dots \otimes {\cal V}^\mathsf{f_6}
\end{equation}
for the space of Hermitian operators.  Hence, we can write any operator as a linear sum over fiducial operators for the inputs and outputs.
\begin{equation}\label{decomposibilityhat}
\hat A_\mathsf{a_1b_2\dots c_3}^\mathsf{d_4e_5\dots f_6} \,\,= {}^{d_4e_5\dots f_6}\!A_{a_1b_2\dots c_3}\,\, \hat {X}_\mathsf{a_1}^{a_1} \hat{X}_\mathsf{b_2}^{b_2} \cdots \hat{X}_\mathsf{c_3}^{c_3} \,\,{}_{d_4}\!\hat{X}^\mathsf{d_4}{}_{e_5}\!\hat{X}^\mathsf{e_5}\cdots {}_{f_6}\!\hat{X}^\mathsf{f_6}
\end{equation}
in symbolic notation, or
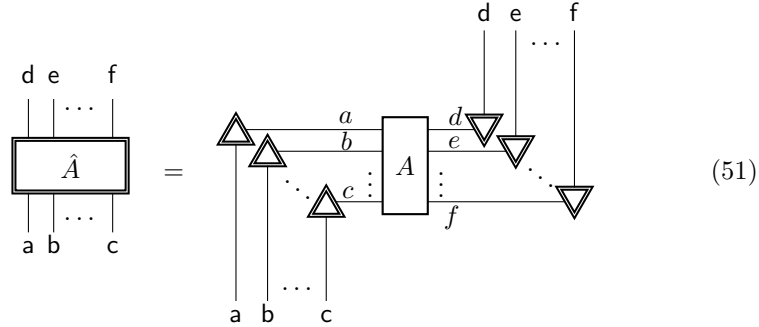
\begin{equation}\label{decomposibilitydiagramhat}
\begin{Diagram}[1.4]{0}{0}
\Dopbox[5]{A}{0,0}
\inwire{A}{1}\Opsymbol{a} \inwire{A}{2.2}\Opsymbol{b} \putlatex[30,15]{\ensuremath{\dots}} \inwire{A}{5}\Opsymbol{c}
\outwire{A}{1}\Opsymbol{d}\outwire{A}{2.2}\Opsymbol{e}\putlatex[30,-50]{\ensuremath{\dots}}   \outwire{A}{5}\Opsymbol{f}
\end{Diagram}
~~~= ~~~
\begin{Diagram}[1.2]{0}{0}
\Duobox[5]{A}{0,0}
\putlatex[-44,-14]{\ensuremath{\vdots}} \putlatex[-135,-20]{\ensuremath{\ddots}}
\putlatex[45,-14]{\ensuremath{\vdots}} \putlatex[170,-5]{\ensuremath{\ddots}}
\dlinkedeffect[0.8]{A}{5}{c}{-3.5}{0}  \duosymbol[-13,-4]{c} \thispoint{cbase}{-3.5,-6}\Opsymbol{c} \wire{cbase}{c}{1}{1}
\dlinkedeffect[0.8]{A}{2.2}{b}{-6.1}{0}\duosymbol[22,-3]{b}
\thispoint{bbase}{-6.1,-6}\Opsymbol{b} \wire{bbase}{b}{1}{1}
\dlinkedeffect[0.8]{A}{1}{a}{-7.5}{0}\duosymbol[41,-1]{a} \thispoint{abase}{-7.5,-6}\Opsymbol{a} \wire{abase}{a}{1}{1}
\placelatex[-20, 16]{-4,-6}{\ensuremath{\dots}}
\dlinkedprep[0.8]{A}{1}{d}{3.5}{0}\duosymbol[4,1]{d}  \thispoint{dbase}{3.5,6}\Opsymbol[0,38]{d} \wire{d}{dbase}{1}{1}
\dlinkedprep[0.8]{A}{2.2}{e}{4.9}{0}\duosymbol[-17,-5]{e}  \thispoint{ebase}{4.9,6}\Opsymbol[0,38]{e} \wire{e}{ebase}{1}{1}
\dlinkedprep[0.8]{A}{5}{f}{7.5}{0}\otherside\duosymbol[-57,-4]{f} \thispoint{fbase}{7.5,6}\Opsymbol[0,38]{f} \wire{f}{fbase}{1}{1}
\placelatex[9, -16]{6,6}{\ensuremath{\dots}}
\end{Diagram}
\end{equation}
in diagrammatic notation.   This is clearly analogous to the assumption of full decomposability for operations.  Notice, however, that (i) it is just a fact about the space of Hermitian operators on a (complex) Hilbert space and (ii) we have equality here rather than equivalence.

\subsection{Combining operator tensors}

Usually, when dealing with operators, much use is made of the tensor product symbol, $\otimes$ and there is, correspondingly, a need to pad expressions with the identity.  We will follow a different route here.   The tensor product of $\hat{A}^\mathsf{a_1}$ and $\hat{B}^\mathsf{b_2}$ is written $\hat{A}^\mathsf{a_1} \hat{B}^\mathsf{b_2}$ (in ${\cal V}^\mathsf{a_1b_2}={\cal V}^\mathsf{a_1}\otimes {\cal V}^\mathsf{b_2}$) rather than $\hat{A}^\mathsf{a_1} \otimes \hat{B}^\mathsf{b_2}$.   More generally we write the tensor product of $\hat C_\mathsf{a_1a_2b_3}^\mathsf{c_3b_5}$ and $\hat D_\mathsf{b_6a_7}^\mathsf{b_8c_9}$ as
\begin{equation}
\hat C_\mathsf{a_1a_2b_3}^\mathsf{c_3b_5} \hat D_\mathsf{b_6a_7}^\mathsf{b_8c_9}
\end{equation}
The order is not important so we could, instead, write
\begin{equation}
\hat D_\mathsf{b_6a_7}^\mathsf{b_8c_9} \hat C_\mathsf{a_1a_2b_3}^\mathsf{c_3b_5}
\end{equation}
since all the relevant information that is normally conveyed by the order in which the tensor product is taken is, here, contained in the integer subscripts on the type labels.

If have a repeated integer this corresponds to taking the product in the Hilbert space corresponding to the repeated index, then taking the partial trace in that space.  To do this in the general case we will use full decomposability.  However, it is worth illustrating this first for the simple case,
\begin{equation}
\hat A^\mathsf{a_1}\hat B_\mathsf{a_1}
~~~\Longleftrightarrow~~~
\begin{Diagram}{0}{-0.5}
\Dopbox[2]{A}{0,0}  \Dopbox[2]{B}{0.5,5} \wire{A}{B}{1.5}{1.5} \opsymbol{a}
\end{Diagram}
\end{equation}
In conventional notation this is equal to $\text{Trace}(\hat A_\mathsf{a_1}\hat D^\mathsf{a_1})$.

Now consider the example
\begin{equation}\label{operatorwiresymbolic}
\hat A_\mathsf{a_1b_2}^\mathsf{b_3c_4}\hat B_\mathsf{a_5b_3}^\mathsf{d_6c_7} =
A_{a_1b_2}^{b_3c_4}\, \hat X^{a_1}_\mathsf{a_1} \hat X^{b_2}_\mathsf{b_2}  \, {}_{b_3}\! \hat X^\mathsf{b_3} \, {}_{c_4} \! \hat X^\mathsf{c_4} \,
B_{a_5b_{3'}}^{d_6c_7}\, \hat X^{a_5}_\mathsf{a_5} \hat X^{b_{3'}}_\mathsf{b_3}  \, {}_{d_6}\! \hat X^\mathsf{d_6} \, {}_{c_7} \! \hat X^\mathsf{c_7}
\end{equation}
Here the $\mathsf{b_3}$ label is repeated so it is implicit that we are taking the trace of the product of ${}_{b_3}\! \hat X^\mathsf{b_3}$ and $\hat X^{b_{3'}}_\mathsf{b_3}$.  We can also write this in diagrammatic notation,
\begin{equation}\label{operatorwirediagramatic}
\begin{Diagram}{0}{-1.2}
\Dopbox[2]{A}{7,0}
\Dopbox[2]{B}{3,7}
\inwire[-5]{A}{1}\Opsymbol{a}
\inwire[5]{A}{2}\Opsymbol{b}
\wire{A}{B}{1}{2}\opsymbol{b}
\outwire[5]{A}{2}\Opsymbol{c}
\inwire[-5]{B}{1}\Opsymbol{a}
\outwire[-5]{B}{1}\Opsymbol{d}
\outwire[5]{B}{2}\Opsymbol{c}
\end{Diagram}
~~~~~~ = ~~~~~~
\begin{Diagram}{0}{-1.5}
\Duobox{A}{0,0}
\dlinkedeffect{A}{1}{1A}{-5}{0} \duosymbol{a}  \inwire{1A}{1} \Opsymbol{a}
\dlinkedeffect{A}{3}{2A}{-3}{0} \duosymbol[0,1]{b}  \inwire{2A}{1} \Opsymbol{b}
\dlinkedprep{A}{1}{A1}{3}{0} \otherside \duosymbol{b}
\dlinkedprep{A}{3}{A2}{5}{0}   \otherside \duosymbol{c}  \outwire{A2}{1} \Opsymbol{c}
\Duobox{B}{-3,10}
\dlinkedeffect{B}{1}{1B}{-8}{0}  \duosymbol{a}  \inwire{1B}{1} \Opsymbol{a}
\dlinkedeffect{B}{3}{2B}{-6}{0} \duosymbol[0,1]{b}
\dlinkedprep{B}{1}{B1}{0}{0}    \otherside \duosymbol{d} \outwire{B1}{1} \Opsymbol{d}
\dlinkedprep{B}{3}{B2}{2}{0}     \otherside\duosymbol{c} \outwire{B2}{1} \Opsymbol{c}
\wire{A1}{2B}{1}{1} \opsymbol{b}
\end{Diagram}
\end{equation}
This Hermitian operator is in
\begin{equation}
{\cal V}_\mathsf{a_1b_2a_5}^\mathsf{c_4d_6c_7} = {\cal V}_\mathsf{a_1}\otimes{\cal V}_\mathsf{b_2} \otimes {\cal V}_\mathsf{a_5} \otimes
{\cal V}^\mathsf{c_4} \otimes {\cal V}^\mathsf{d_6} \otimes {\cal V}^\mathsf{c_7}
\end{equation}
Note that we have taken the partial trace over the space associated with $\mathsf{b_3}$.  We will call the product in (\ref{operatorwiresymbolic}, \ref{operatorwirediagramatic}) the \emph{circuit trace} since, once we expand in terms of fiducials, we are taking the trace over a fiducial circuit consisting of a fiducial preparation operator and a fiducial result operator.

More generally, we can consider expressions such as
\begin{equation}
\hat{A}^\mathsf{a_1b_2}\hat{B}_\mathsf{b_2}^\mathsf{c_3a_4}\hat{C}_\mathsf{a_5c_3a_4}^\mathsf{b_6} \in {\cal V}_\mathsf{a_5}^\mathsf{a_1b_6}
\end{equation}
Here we have three repeated labels ($\mathsf{b_2}$, $\mathsf{c_3}$, and $\mathsf{a_4}$).  Once expanded in terms of fiducial operators each of these repeated labels will correspond to a fiducial operator circuit.

\subsection{The operator hopping metric}

Whenever we wire together operator tensors and then insert the fully decomposed form the \emph{operator hopping metric} will appear.  This is
\begin{equation}\label{operatorhopping}
{}_{a'_1}\! g^{a_1} := {}_{a'_1}\!\hat{X}^\mathsf{a_1}\hat{X}_\mathsf{a_1}^{a_1} ~~~~~\Leftrightarrow ~~~~~
\begin{Diagram}{0}{-0.1}
\bbmetric{g}{0,0}\duosymbol{a}
\end{Diagram}
~:=~
\begin{Diagram}{0}{-0.6}
\dfidprep{X2}{4,0}
\dfideffect{X}{4,4}
\wire{X2}{X}{1}{1}\opsymbol{a}
\inblack{X2}{1} \Duosymbol{a}
\outblack{X}{1}\Duosymbol{a}
\end{Diagram}
\end{equation}
Strictly, we should use a different symbol than for the hopping metric deriving from fiducial operations.  However, we will be seeking to set them equal in the next section so we will use the same symbol.

\subsection{General operator circuit}

We can use full decomposability of operators to convert an operator circuit into a duotensor calculation.  For example,
\begin{flalign*}
&\begin{Diagram}{0}{-2}
\Dopbox{A}{0,0} \Dopbox[2]{C}{4,6} \Dopbox[2]{B}{-3,10} \Dopbox[2]{D}{1,15}
\wire{A}{B}{1}{1}\opsymbol{a} \wire{A}{C}{2}{1}\opsymbol[0,6]{c} \wire{A}{C}{3}{2}\otherside\opsymbol{a} \wire{C}{B}{1}{2}\opsymbol{a}
\wire{C}{D}{2}{2}\otherside\opsymbol[4,0]{d} \wire{B}{D}{1.5}{1}\opsymbol[0,6]{b}
\end{Diagram}
~=~
\begin{Diagram}{0}{-3}
\begin{move}{-7,5}
\Duobox{A}{0,0}
\dlinkedprep[0.7]{A}{1}{A1}{3}{0}\duosymbol[-3,-6]{\scriptstyle a}
\dlinkedprep[0.7]{A}{2}{A2}{4.5}{0}\duosymbol[-25,-6]{\scriptstyle c}
\dlinkedprep[0.7]{A}{3}{A3}{6}{0}\duosymbol[-44,-6]{\scriptstyle a}
\end{move}
\begin{move}{10,15}
\Duobox[2]{B}{0,0}
\dlinkedeffect[0.7]{B}{1}{1B}{-4.5}{0}\duosymbol[25,-6]{\scriptstyle a}
\dlinkedeffect[0.7]{B}{2}{2B}{-3}{0}\duosymbol[3,-6]{\scriptstyle a}
\dlinkedprep[0.7]{B}{1.5}{B15}{3}{0} \duosymbol[-3,-3]{\scriptstyle b}
\end{move}
\begin{move}{4,8}
\Duobox[2]{C}{0,0}
\dlinkedeffect[0.7]{C}{1}{1C}{-4.5}{0}\duosymbol[25,-6]{\scriptstyle c}
\dlinkedeffect[0.7]{C}{2}{2C}{-3}{0}\duosymbol[3,-6]{\scriptstyle a}
\dlinkedprep[0.7]{C}{1}{C1}{3}{0} \duosymbol[-3,-6]{\scriptstyle a}
\dlinkedprep[0.7]{C}{2}{C2}{4.5}{0} \otherside\duosymbol[-25,3]{\scriptstyle d}
\end{move}
\begin{move}{19,18}
\Duobox[2]{D}{0,0}
\dlinkedeffect[0.7]{D}{1}{1D}{-4.5}{0}\duosymbol[25,-6]{\scriptstyle b}
\dlinkedeffect[0.7]{D}{2}{2D}{-3}{0}\otherside\duosymbol[3,3]{\scriptstyle d}
\end{move}
\wire{A1}{1B}{1}{1}\opsymbol{a} \wire[0.4]{A2}{1C}{1}{1}\opsymbol{c} \wire[0.4]{A3}{2C}{1}{1}\otherside\opsymbol{a}
\wire{C1}{2B}{1}{1}\opsymbol{a} \wire{C2}{2D}{1}{1}\otherside\opsymbol{d} \wire[0.4]{B15}{1D}{1}{1}\opsymbol[-3,0]{b}
\end{Diagram}
\\
&=~
\begin{Diagram}[1]{0}{0}
\Duobox{A}{-3,0} \Duobox[2]{B}{11,3} \Duobox[2]{C}{6,-4}  \Duobox[2]{D}{18,-1}
\linkwbbw[0.6]{A}{B}{1}{1}\duosymbol{a} \linkwbbw[0.6]{A}{C}{2}{1}\duosymbol{c} \linkwbbw[0.6]{A}{C}{3}{2}\otherside\duosymbol{a}
\linkwbbw[0.6]{C}{B}{1}{2}\duosymbol{a} \linkwbbw[0.6]{C}{D}{2}{2}\duosymbol{d} \linkwbbw[0.6]{B}{D}{1.5}{1}\duosymbol{b}
\end{Diagram}
~=~
\begin{Diagram}{0}{0}
\Duobox{A}{0,0} \Duobox[2]{B}{10,3} \Duobox[2]{C}{6,-4}  \Duobox[2]{D}{15,-1}
\link{A}{B}{1}{1}\duosymbol{a} \link{A}{C}{2}{1}\duosymbol{c} \link{A}{C}{3}{2}\otherside\duosymbol{a}
\link{C}{B}{1}{2}\duosymbol{a} \link{C}{D}{2}{2}\duosymbol{d} \link{B}{D}{1.5}{1}\duosymbol{b}
\end{Diagram}
\end{flalign*}
This is very similar to the operation case discussed in Sec.\ \ref{generalcircuits} where we used full decomposability of operations, the main difference being that we have equality rather than equivalence in the first and second step.  This is because we have equality in the full decomposability of operators (\ref{decomposibilitydiagramhat}), but only equivalence in the full decomposability of operations (\ref{assumption2diagram}).

\section{Operator tensor formulation of QT}

\subsection{Correspondence}

We will say that operations correspond to operators if there is a mapping from operations, $\mathsf{A}_\mathsf{a_1b_2\dots c_3}^\mathsf{d_4e_5\dots f_6}$, to operators, $\hat{A}_\mathsf{a_1b_2\dots c_3}^\mathsf{d_4e_5\dots f_6}$, such that the probability for \emph{any} circuit comprised of operations is equal to the circuit operator obtained under this mapping.  If operations correspond to operators then, for example
\begin{equation}
\text{Prob}(\mathsf{A}^\mathsf{a_1b_2}\mathsf{B}_\mathsf{b_2}^\mathsf{c_3a_4}\mathsf{C}_\mathsf{a_1c_3a_4})
=\hat{A}^\mathsf{a_1b_2}\hat{B}_\mathsf{b_2}^\mathsf{c_3a_4}\hat{C}_\mathsf{a_1c_3a_4}
\end{equation}
We can write the same example in diagrammatic notation
\begin{equation}\label{probtraceexample}
\text{Prob}\left(~
\begin{Diagram}{0}{-1.4}
\Opbox[2]{A}{0,0} \Opbox[2]{B}{2,4} \Opbox{C}{-1,10}
\wire{A}{C}{1}{1} \opsymbol{a} \wire{A}{B}{2}{1.5}\opsymbol[-3,1]{b} \wire{B}{C}{1}{2} \opsymbol{c} \wire{B}{C}{2}{3}\otherside\opsymbol{a}
\end{Diagram}
~\right)
~=~
\begin{Diagram}{0}{-1.4}
\Dopbox[2]{A}{0,0} \Dopbox[2]{B}{2,4} \Dopbox{C}{-1,10}
\wire{A}{C}{1}{1} \opsymbol{a} \wire{A}{B}{2}{1.5}\opsymbol[-3,1]{b} \wire{B}{C}{1}{2} \opsymbol{c} \wire{B}{C}{2}{3}\otherside\opsymbol{a}
\end{Diagram}
\end{equation}
In fact, if we set up a correspondence from fiducial operations to fiducial operators then a correspondence from general operations to general operators follows.  To see this assume we have found a correspondence
\begin{equation}
\mathsf{X}_\mathsf{a_1}^{a_1} \rightarrow \hat X_\mathsf{a_1}^{a_1} ~~~~\text{and}~~~~ {}_{a_1}\!\mathsf{X}^\mathsf{a_1} \rightarrow {}_{a_1}\!\hat{X}^\mathsf{a_1}
\end{equation}
for $a_1=1$ to $K_\mathsf{a}$ for all types $\mathsf{a}$.  Since we have correspondence, it follows that
\begin{equation}
\text{Prob}( {}_{a_1}\!\mathsf{X}^\mathsf{a_1} \mathsf{X}_\mathsf{a_1}^{a'_1}) = {}_{a_1}\!\hat{X}^\mathsf{a_1} \hat{X}_\mathsf{a_1}^{a'_1}
\end{equation}
In other words, the hopping metric associated with the operations is equal to the hopping metric associated with the operators for each type.  Then, in general, we can say that the operation
\begin{equation}\label{operationcorrespondsto}
\mathsf{A_{a_1b_2\dots c_3}^{d_4e_5\dots f_6}} \,\,\equiv {}^{d_4e_5\dots f_6}\!A_{a_1b_2\dots c_3}\,\, \mathsf{X}_\mathsf{a_1}^{a_1} \mathsf{X}_\mathsf{b_2}^{b_2} \cdots \mathsf{X}_\mathsf{c_3}^{c_3} \,\,{}_{d_4}\!\mathsf{X}^\mathsf{d_4}{}_{e_5}\!\mathsf{X}^\mathsf{e_5}\cdots {}_{f_6}\!\mathsf{X}^\mathsf{f_6}
\end{equation}
corresponds to the operator
\begin{equation}\label{operatorcorrespondsto}
\hat A_\mathsf{a_1b_2\dots c_3}^\mathsf{d_4e_5\dots f_6} \,\,= {}^{d_4e_5\dots f_6}\!A_{a_1b_2\dots c_3}\,\, \hat {X}_\mathsf{a_1}^{a_1} \hat{X}_\mathsf{b_2}^{b_2} \cdots \hat{X}_\mathsf{c_3}^{c_3} \,\,{}_{d_4}\!\hat{X}^\mathsf{d_4}{}_{e_5}\!\hat{X}^\mathsf{e_5}\cdots {}_{f_6}\!\hat{X}^\mathsf{f_6}
\end{equation}
Note we have the same duotensor in each expansion. We can illustrate that this gives us operation-operator correspondence by means of an example.  Thus, we have \begin{equation}
\text{Prob}\left(~
\begin{Diagram}{0}{-1.4}
\Opbox[2]{A}{0,0} \Opbox[2]{B}{2,4} \Opbox{C}{-1,10}
\wire{A}{C}{1}{1} \opsymbol{a} \wire{A}{B}{2}{1.5}\opsymbol[-3,1]{b} \wire{B}{C}{1}{2} \opsymbol{c} \wire{B}{C}{2}{3}\otherside\opsymbol{a}
\end{Diagram}
~\right)
~=~
\begin{Diagram}{0}{0}
\Duobox[2]{A}{0,0} \Duobox[2]{B}{7,-3} \Duobox{C}{16,1.5}
\linkwbbw{A}{B}{2}{1.5} \duosymbol{b}
\linkwbbw{A}{C}{1}{1}   \duosymbol{a}
\linkwbbw{B}{C}{1}{2}   \duosymbol{c}
\linkwbbw{B}{C}{2}{3}  \otherside \duosymbol{a}
\end{Diagram}
~=~
\begin{Diagram}{0}{-1.4}
\Dopbox[2]{A}{0,0} \Dopbox[2]{B}{2,4} \Dopbox{C}{-1,10}
\wire{A}{C}{1}{1} \opsymbol{a} \wire{A}{B}{2}{1.5}\opsymbol[-3,1]{b} \wire{B}{C}{1}{2} \opsymbol{c} \wire{B}{C}{2}{3}\otherside\opsymbol{a}
\end{Diagram}
\end{equation}
Here we are using full decomposability of operations to get from the first to the second expression, and full decomposability of operators to get from the third to the second expression.  The fact that we have equal hopping metrics means that we have equality of the first and third expressions (as in (\ref{probtraceexample})).    Clearly this will work for any circuit.

\subsection{Experimentally determining operators}\label{Experimentaldeterminationsection}

Here we will discuss how to determine experimentally the operator associated with an operation.  First we need to set up fiducial sets of preparations, ${}_{a_1}\! \mathsf{X}^\mathsf{a_1}$, and fiducial sets of results, $\mathsf{X}_\mathsf{a_1}^{a_1}$.  This can be done either by having a knowledge of the physics (for example, we know that for a spin half particle, results corresponding to spin along the $+x$, $+y$, $+z$, and $-z$ directions will suffice), or through exhaustive tomography on all the preparations and results associated with the given system type followed by linear analysis of the results to pick out fiducial sets.  Having established these fiducial sets, we can measure the hopping metric by experimentally determining $\text{Prob}({}_{a'_1}\! \mathsf{X}^\mathsf{a_1} \mathsf{X}_\mathsf{a_1}^{a_1})$.  Next, by means of appropriate linear algebra, we find sets of fiducial operator preparations and results that have the same hopping metric, i.e.\
\begin{equation}
\text{Prob}({}_{a'_1}\! \hat{X}^\mathsf{a_1} \hat{X}_\mathsf{a_1}^{a_1}) =
\text{Prob}({}_{a'_1}\! \mathsf{X}^\mathsf{a_1} \mathsf{X}_\mathsf{a_1}^{a_1})
\end{equation}
Having obtained the hopping metric and the fiducial preparation and result operators we can determine the operator associated with an operation as follows.
\begin{enumerate}
\item First we perform local process tomography placing fiducial preparations on all the inputs and fiducial results on all the outputs.  The set of probabilities so obtained is equal to the duotensor with all black dots (see Sec.\ \ref{duotensorssubsection}).
\item Next, obtain the inverse of hopping metric and use this to convert the duotensor to the form having all white dots.
\item Finally, calculate the operator tensor by weighting fiducial operators with this all white dots duotensor as in (\ref{decomposibilityhat}) or  (\ref{decomposibilitydiagramhat}).
\end{enumerate}

\subsection{Physicality of operators}

We do not expect all operators to correspond to operations.   Thus, the set of operators that are allowed will be restricted in some way.  We assume that, for every type, the set of operators that are associated with operations contains at least the following.
\begin{enumerate}
\item Every rank one projection preparation operator, $\hat A^\mathsf{a_1}$.
\item Every rank one projection result operator, $\hat C_\mathsf{a_1}$.
\item The identity result operator, $\hat I_\mathsf{a_1}$.
\end{enumerate}
Recall that a type, $\mathsf{a}$, may correspond to a composite system.
The identity result operator has the property that it gives 1 when applied to any rank one projection preparation operator (e.g. $\hat A^\mathsf{a_1} \hat I_\mathsf{a_1}=1$). Thus, it must correspond to the deterministic result (where the set of outcomes is equal to the set of all possible outcomes).

We define
\begin{quote}
\item {\bf Physical operators.} An operator, $\hat B_\mathsf{a_1}^\mathsf{b_2}$, is said to be \emph{physical} if
\begin{equation}\label{positiveoperatorcondition}
0\leq \hat A^\mathsf{a_1g_3} \hat B_\mathsf{a_1}^\mathsf{b_2} \hat C_\mathsf{b_2g_3}
\end{equation}
and
\begin{equation}\label{traceoperatorcondition}
\hat A^\mathsf{a_1g_3} \hat B_\mathsf{a_1}^\mathsf{b_2}\hat I_\mathsf{b_2g_3} \leq 1
\end{equation}
for all rank one projection operators $\hat A^\mathsf{a_1g_7}$ and $C_\mathsf{d_4g_7}$ and for all types $\mathsf g$.
\end{quote}
This means that any circuit built out of the given operator sandwiched between the operators we have already admitted (rank one projection preparation and result operators and the identity result operator) will be between 0 and 1 as is necessary for the probability interpretation.

We define the \emph{input transpose} of an operator as the partial transpose in the input space with respect to some given basis.  We denote the input transpose of $\hat B_\mathsf{a_1}^\mathsf{b_2}$ by $\hat B_\mathsf{a^T_1}^\mathsf{b_2}$.  One way to find the input transpose is to write the operator in fully decomposed form and then take the transpose of the fiducial result operators (as these correspond to the inputs).  We note that, if an operator is positive after taking the input transpose with respect one basis then the input transpose will be positive when taken with respect to any other basis.  We can define the \emph{output transpose} in a similar way.  More generally, we can consider expressions such as $\hat A_\mathsf{a^T_1b_2}^\mathsf{c_3d^T_4}$ where we take the partial transpose over some of the spaces.  We note a useful result.  Namely, that if we take the partial transpose in any space over which the circuit trace is taken then the expression remains unchanged.  For example
\begin{equation}
\hat A_\mathsf{a_1b_2}^\mathsf{c_3d_4} \hat B^\mathsf{b_2c_5}_\mathsf{a_6} = \hat A_\mathsf{a_1b^T_2}^\mathsf{c_3d_4} \hat B^\mathsf{b^T_2c_5}_\mathsf{a_6}
\end{equation}
This is simple to prove (see Sec.\ 8.3 of \cite{hardy2011reformulating}).

We will now prove the following theorem.
\begin{quote}
\item {\bf Physicality theorem.} An operator, $\hat B_\mathsf{a_1}^\mathsf{b_2}$,  is physical if and only if (a) it has positive input transpose i.e. $\hat B_\mathsf{a^T_1}^\mathsf{b_2} \geq 0$ and (b) its output trace is less than or equal to the identity, i.e. $\hat B_\mathsf{a_1}^\mathsf{b_2} \hat I_\mathsf{b_2} \leq I_\mathsf{a_1}$.
\end{quote}
To prove this note that we can write
\begin{equation}
\hat A^\mathsf{a_1g_3} = |A^\mathsf{a_1g_3}\rangle \langle A^\mathsf{a_1g_3} |  ~~~~~~~
\hat C_\mathsf{b_2g_3} = |C_\mathsf{b_2g_3} \rangle \langle C_\mathsf{b_2g_3}|
\end{equation}
We will consider taking the output transpose with respect to the basis $|V^\mathsf{a_1}[n]\rangle$ ($n=1$ to $N_\mathsf{a}$).  We can write, in Schmidt decomposed form,
\begin{equation}
|A^\mathsf{a_1g_3}\rangle = \sum_n c_n|V^\mathsf{a_1}[n]\rangle|W^\mathsf{g_3}[n]\rangle
\end{equation}
where the states $|\hat W^\mathsf{g}[n]\rangle$ constitute an orthonormal basis set for $\mathcal{H}^\mathsf{g_3}$ allowing this Schmidt decomposition. We can write
\begin{equation}
\langle C_\mathsf{b_2g_3}| = \sum_n  \langle E_\mathsf{b_2}[n]|\langle W_\mathsf{g_3}[n]|
\end{equation}
Then
\begin{equation}
\hat A^\mathsf{a_1g_3} \hat C_\mathsf{b_2g_3} = \left(\sum_m c^*_m \langle V^\mathsf{a_1}[m]| \otimes | E_\mathsf{b_2}[m] \rangle  \otimes\left( c_n \sum_n |V^\mathsf{a_1}[n]\rangle \otimes \langle E_\mathsf{b_2}[n] | \right)
\right)
\end{equation}
where we have used the fact that we are taking the partial trace over the space associated with $\mathsf{g_3}$ (we have also inserted the $\otimes$ symbol for clarity).  The output transpose of this (in the $|V^\mathsf{a_1}[n]\rangle$ basis) is
\begin{equation}\label{aftertranspose}
\hat A^\mathsf{a^T_1g_3} \hat C_\mathsf{b_2g_3} = \left( \sum_m c^*_m |V^\mathsf{a_1}[m]\rangle \otimes | E_\mathsf{b_2}[m] \rangle \right)\otimes \left(\sum_n  c_n \langle V^\mathsf{a_1}[n]| \otimes \langle E_\mathsf{b_2}[n] |\right)
\end{equation}
This is proportional to a rank one projection operator.  We are free to choose $c_n$ and $|E_\mathsf{b_2}[n]\rangle$ as we like (subject only to normalization constraints) and, hence, we can make this proportional to any rank one operator.  Now we note that
\begin{equation}\label{ABCcircuit}
\hat A^\mathsf{a_1g_3} \hat B_\mathsf{a_1}^\mathsf{b_2} \hat C_\mathsf{b_2g_3} = \hat A^\mathsf{a^T_1g_3} \hat C_\mathsf{b_2g_3} \hat B_\mathsf{a^T_1}^\mathsf{b_2}
\end{equation}
since, as we noted above, such expressions are invariant under taking the transpose in any given space over which the partial trace is taken.   Since the expression in  (\ref{aftertranspose}) can be made to be proportional to a rank one operator and since the expression in (\ref{ABCcircuit}) must be non-negative by (\ref{positiveoperatorcondition}), it follows that $\hat B_\mathsf{a^T_1}^\mathsf{b_2}$ must be positive.  This proves the necessity of the first point.  The necessity of the second point follows immediately from (\ref{traceoperatorcondition}), the fact that $\hat I_\mathsf{b_2g_3}=\hat I_\mathsf{b_2}\hat I_\mathsf{g_3}$, and the fact that $\hat A^\mathsf{a_1g_3}$ can correspond to any rank one projection operator.  It is also clear, by these methods, that sufficiency follows.   Hence the theorem is proven.

We can easily see that any operator circuit comprised only of physical operators will be equal to a number between 0 and 1 (as required for the probability interpretation).   To illustrate this consider the example,
\begin{equation}
\hat{A}^\mathsf{a_1b_2}\hat{B}_\mathsf{b_2}^\mathsf{c_3a_4}\hat{C}_\mathsf{a_1c_3a_4}
\end{equation}
(this example is shown diagrammatically on the left of (\ref{probtraceexample})).
We can foliate this circuit
\begin{equation}\label{simplefoliatedexample}
[\hat{A}^\mathsf{a_1b_2}   ] [ \hat I_\mathsf{a_1}^\mathsf{a_5} \hat{B}_\mathsf{b_2}^\mathsf{c_3a_4}] [\hat C_\mathsf{a_5c_3a_4} ]
\end{equation}
where we have indicated time steps (and hence hypersurfaces of the foliation) by square brackets.  We have had to insert the identity transformation, $I_\mathsf{a_1}^\mathsf{a_5}$, to allow this foliation as two hypersurfaces intercept the given wire.  Now, we take the partial transpose on the spaces intercepted by \emph{alternate} hypersurfaces
\begin{equation}\label{simplefoliatedwithTsexample}
[\hat{A}^\mathsf{a^T_1b^T_2}   ] [ \hat I_\mathsf{a^T_1}^\mathsf{a_5}\hat{B}_\mathsf{b^T_2}^\mathsf{c_3a_4} ] [ C_\mathsf{a_5c_3a_4} ]
\end{equation}
Now any given operator has either had the input or output transpose taken and, since all operators in (\ref{simplefoliatedexample}) are physical, all the operators in (\ref{simplefoliatedwithTsexample}) are now positive (positivity of input transpose is equivalent to positivity of output transpose).  Hence, the expression must be positive.  This will work for any circuit, no matter how many hypersurfaces are needed to foliate it, as long as we take the transpose on alternate hypersurfaces.   That the operator circuit is less or equal than 1 follows by by proceeding backwards through foliation.   Thus, $[\hat C_\mathsf{a_5c_3a_4}]$ is less than $\hat I_\mathsf{a_5c_3a_4}$.  Hence, $[ \hat I_\mathsf{a_1}^\mathsf{a_5} \hat{B}_\mathsf{b_2}^\mathsf{c_3a_4}] [\hat C_\mathsf{a_5c_3a_4} ]$ is less than $\hat I_\mathsf{a_1b_2}$.  Hence, the whole thing is less or equal to 1.  It is interesting that we invoke the possibility of foliating the circuit here.  A circuit can be foliated if and only if there are no closed loops (see, for example, \cite{hardy2009foliable}).  Hence, there is a deep connection between such operator circuits behaving like probabilities (being bounded by 0 and 1) and the lack of such closed loops (which would correspond to closed time-like curves).

We define
\begin{quote}
{\bf A complete set of physical operators}, $\{\hat B_\mathsf{a_1}^\mathsf{b_2}[l]: l=1 ~\text{to}~ L\}$, is one in which each operator, $\hat B_\mathsf{a_1}^\mathsf{b_2}[l]$ has positive input transpose and, further,
\begin{equation}\label{completephysicalcond}
\sum_{l=1}^L  \hat B_\mathsf{a_1}^\mathsf{b_2}[l] \hat I_\mathsf{b_2}  = \hat I_\mathsf{a_1}
\end{equation}
\end{quote}
The operators in a complete set of physical physical operators must each be physical as (\ref{completephysicalcond}) implies $\hat B_\mathsf{a_1}^\mathsf{b_2}[l]  \hat I_\mathsf{b_2} \leq  \hat I_\mathsf{a_1}$ (using the fact that $\hat B_\mathsf{a_1}^\mathsf{b_2}[l]  \hat I_\mathsf{b_2} = \hat B_\mathsf{a_1}^\mathsf{b^T_2}[l]  \hat I_\mathsf{b^T_2} \geq 0$ ).  Further, (\ref{completephysicalcond}) is equivalent to the condition that
\begin{equation}
\sum_{l=1}^L  \hat A^\mathsf{a_1g_3}  \hat B_\mathsf{a_1}^\mathsf{b_2}[l] \hat I_\mathsf{b_2g_3} = 1
\end{equation}
for any positive operator, $\hat A^\mathsf{a_1g_3}$, having trace equal to one.

\subsection{Mathematical axioms for quantum theory}

We will now give axioms for quantum theory in the operator tensor formulation.
\begin{quote}\index{axioms}
{\bf QUANTUM THEORY.}  The following two mathematical axioms specify quantum theory:
\begin{description}
\item[Axiom 1] \index{axioms!\textbf{Axiom 1}}Operations correspond to operators.
\item[Axiom 2] \index{axioms!\textbf{Axiom 2}}Every complete set of physical operators corresponds to a complete set of operations.
\end{description}
The operators here are understood to act on a complex Hilbert space.
\end{quote}
Recall that a complete set of operations is one in which each member is associated with the same apparatus use with the same settings but having disjoint outcome sets whose union is the set of all outcomes for this apparatus use.  The first axiom guarantees that we can calculate the probability of a circuit (comprised of operations) by replacing each operation by the corresponding operator.  The second axiom tells us that every permissible set of operators (i.e.\ every complete set of physical operators) actually corresponds to a realizable complete set of operations.  A more succinct statement is the following
\begin{quote}
{\bf QUANTUM THEORY:} Every complete set of positive operators corresponds to a complete set of operations and vice versa.
\end{quote}
The vice versa part of this implies Axiom 1 though, taken at face value, it is a little stronger than axiom 1.

These axioms are motivated by assuming that certain operators are allowed (rank one projection operators for preparations and results, and the identity operator for results) and seeing what the largest class of operators consistent with this situation is.   We get an alternative formulation of quantum theory in this way which is easily shown to be equivalent to the usual formulation as discussed in Sec.\ \ref{oldformulation} (see \cite{hardy2011reformulating} for more details).

\subsection{Formalism locality}

So far we have considered how to calculate the probability for a circuit.  However, we may be interested in doing calculations for some part of a circuit without doing a calculation for the whole circuit.  An extreme example would be where we model the entire universe by means of a circuit (let us assume that the universe is finite such that the circuit has a finite number of operations).  In this case we would, for many applications, need to be able to make predictions concerning only a part of the universe.   In the circuit framework an arbitrary spacetime region corresponds to a fragment (more accurately, it corresponds to the fragments having some given causal structure that can be placed in a corresponding \lq\lq hole" in the bigger circuit).  These considerations motivate the following definition
\begin{quote}
{\bf Formalism locality.}  A formalism for a physical theory is said to have the property of \lq\lq formalism locality" if we can do calculations
pertaining to any region of spacetime employing only mathematical objects associated with that region.
\end{quote}
Formalism locality is a property of the way a physical theory is formulated rather than, necessarily, being a property of the theory itself.  We will show that the operator tensor formulation has this property.  Thus, consider a set of fragment, $\mathsf{A}_{a_1b_2\dots c_3}^{d_4e_5\dots g_6}[n]$ all having the same causal structure (see Sec.\ \ref{wiresandfragments}). The most general thing we can consider calculating for this situation is the probability ratio
\begin{equation}\label{probratio}
\frac{\text{Prob}(\mathsf{A}_{a_1b_2\dots c_3}^{d_4e_5\dots g_6}[m])}{\text{Prob}(\mathsf{A}_{a_1b_2\dots c_3}^{d_4e_5\dots g_6}[n])}
\end{equation}
What does this mean?  In particular, in most cases we expect this probability to depend on what is done outside the given spacetime region - in other words we expect that it will, in most cases, depend on what the bigger circuit is.  However, this may not always be the case. Thus, what we are seeking is (a) some mathematical condition telling us when this probability ratio is well defined (independent of what happens outside the given spacetime region) and, (b) in such situations, we also want to know what this probability ratio is equal to.

If we complete the fragments into a circuit in the same way then we have
\begin{equation}\label{probratioexpanded}
\frac{\text{Prob}(\mathsf{A}_{a_1b_2\dots c_3}^{d_4e_5\dots g_6}[m]  \mathsf{E}^{a_1b_2\dots c_3}_{d_4e_5\dots g_6}) }
     {\text{Prob}(\mathsf{A}_{a_1b_2\dots c_3}^{d_4e_5\dots g_6}[n]  \mathsf{E}^{a_1b_2\dots c_3}_{d_4e_5\dots g_6}) }
     =
\frac{\hat{A}_\mathsf{a_1b_2\dots c_3}^\mathsf{d_4e_5\dots g_6}[m]  \hat{E}^\mathsf{a_1b_2\dots c_3}_\mathsf{d_4e_5\dots g_6} }
     {\hat{A}_\mathsf{a_1b_2\dots c_3}^\mathsf{d_4e_5\dots g_6}[n]  \hat{E}^\mathsf{a_1b_2\dots c_3}_\mathsf{d_4e_5\dots g_6}}
\end{equation}
The only way for the probability ratio in (\ref{probratio}) sense is if the expression in (\ref{probratioexpanded}) is independent of $\hat{E}^\mathsf{a_1b_2\dots c_3}_\mathsf{d_4e_5\dots g_6}$.  But $\hat{E}^\mathsf{a_1b_2\dots c_3}_\mathsf{d_4e_5\dots g_6}$ can span the space of possible Hermitian operators and hence, the only way to have this independence is if $\hat{A}_\mathsf{a_1b_2\dots c_3}^\mathsf{d_4e_5\dots g_6}[m]$ and $\hat{A}_\mathsf{a_1b_2\dots c_3}^\mathsf{d_4e_5\dots g_6}[n]$ are proportional
\begin{equation}
\hat{A}_\mathsf{a_1b_2\dots c_3}^\mathsf{d_4e_5\dots g_6}[m]= r\hat{A}_\mathsf{a_1b_2\dots c_3}^\mathsf{d_4e_5\dots g_6}[n]
\end{equation}
The probability ratio is then equal to the proportionality constant, $r$.   Hence we have formalism locality.  Standard formulations of quantum theory do not have this property.

\subsection{Transforming operator tensors}

We can transform an operator tensor in a similar way to the way we would a tensor.  We must be sure that any operator circuit remains unchanged under such a transformation (operator circuits are the equivalent of scalars for tensors).   Thus, for each system type $\mathsf a$, we can write
\begin{equation}
\hat X_\mathsf{a'_1}^{a_1} =  \hat T_\mathsf{a'_1}^\mathsf{a_1}  \hat X_\mathsf{a_1}^{a_1}
~~~~~~~~~~~
{}_{a_1}\! \hat X^\mathsf{a'_1}  = \hat T_\mathsf{a_1}^\mathsf{a'_1} {}_{a_1}\!\hat X^\mathsf{a_1}
\end{equation}
where, importantly, we require
\begin{equation}
\hat T_\mathsf{a_1}^\mathsf{a'_1}\hat T_\mathsf{a'_1}^\mathsf{a_2} = \hat I_\mathsf{a_1}^\mathsf{a_2}
\end{equation}
where $\hat I_\mathsf{a_1}^\mathsf{a_2}$ corresponds to the identity transformation having the properties that $\hat I_\mathsf{a_1}^\mathsf{a_2} \hat C_\mathsf{a_3}^\mathsf{a_1} = \hat C_\mathsf{a_3}^\mathsf{a_2}$ and $\hat I_\mathsf{a_1}^\mathsf{a_2} \hat C_\mathsf{a_2}^\mathsf{a_4} = \hat C_\mathsf{a_1}^\mathsf{a_4}$ for any operator $\hat C_\mathsf{a_3}^\mathsf{a_4}$.  Hence we have
\begin{equation}
{}_{a_1}\! \hat X^\mathsf{a'_1} \hat X_\mathsf{a'_1}^{a_1} ={}_{a_1}\! \hat X^\mathsf{a_1} \hat X_\mathsf{a_1}^{a_1}
\end{equation}
as this is an operator circuit (it has no open wires).  This expression is, in fact, the hopping metric (it is a duotensor) and consequently it follows that under such transformations, any operator circuit will remain unchanged.

What transformations, $\hat T_\mathsf{a'_1}^\mathsf{a_1}$, can we allow?  If we are only interested in leaving operator circuits invariant then any invertible transformation will do.  However, we may also be interested in being sure that physical operators transform to physical operators.  We started by assuming that the set of rank one projectors correspond to preparations.  It follows from the fact that preparation operators must be positive and have trace less than or equal to one that such rank one projectors represent pure states, and that all pure states are represented by such rank one projectors.  Therefore, if we wish to preserve physicality under a transformation, rank one projector preparations must be transformed to rank one projector preparations.  By Wigner's theorem \cite{wigner1959group} we know this is only true if we have transformations corresponding to unitary or anti-unitary maps on the underlying Hilbert space (see Peres's book \cite{peres1993quantum} for a nice discussion of Wigner's theorem).  Consider an operator tensor, $\hat T_\mathsf{a'_1}^\mathsf{a_1}$, corresponding to a unitary map
\begin{equation}
\hat T_\mathsf{a_1}^\mathsf{a'_1} \hat A^\mathsf{a_1} = U \hat A^\mathsf{a_1} U^\dagger
\end{equation}
(Here we have not attempted to balance the subscripts and superscripts as the notation on the left hand side is foreign to the operator tensor notation generally adopted in this paper.)  Such transformations will clearly preserve physicality of a general operator tensor.  What about the anti-unitary case? Consider a preparation operator $\hat A^\mathsf{a_1b_2}$ corresponding to a pure entangled state.  Any anti-unitary map can be written as complex conjugation followed by a unitary map.  At the level of the space $\cal{V}^\mathsf{a_1}$, this corresponds to a transposition transformation followed by a transformation corresponding to a unitary map.  Thus, the essential part of the transformation corresponding to an anti-unitary map when applied to one system is partial transposition for that system.  If we perform a partial transposition for the system of type $\mathsf{a}$ but perform no transformation for the system of type $\mathsf{b}$ then, after the transformation, the preparation operator will no longer be physical in general (this follows from the work of Peres \cite{peres1996separability}).  Hence, if we include transformations corresponding to anti-unitary maps on the underlying Hilbert space, then we will not preserve physicallity.

With standard tensors we can restrict ourselves to orthogonal transformations such that the coordinate system remains orthogonal, or we can consider general invertible transformations such that we consider general coordinate systems (as in General Relativity).   Similarly, with operator tensors we can restrict ourselves to unitary transformations such that the operators remain physical, or we can relax this and consider general invertible transformation and then the operators will become non-physical.  For the purposes of this paper it is better to stick with physicallity preserving transformations so the axioms above apply after a transformation.  However, it is possible that for some other application the freedom to have general transformations (such that we go to non-physical operators) we be important just as the freedom of general coordinate transformations (such that we go to non orthogonal coordinate systems) was important in formulating General Relativity.

\section{Discussion}

\subsection{Time symmetry and asymmetry}

The standard formulation of quantum theory requires that we foliate a circuit so we can evolve a state in time.  This means adding extra structure to a circuit that is not part of the physics.  The operator tensor formulation does not do this. Rather, it supplies a formulation of quantum theory in which the calculation looks the same as the description of the circuit.  Operations correspond to operators. An operator, $\hat A_\mathsf{a_1b_2}^\mathsf{c_3d_4}$ for example, must satisfy two conditions (i) that it has positive input transpose ($\hat A_\mathsf{a^T_1b^T_2}^\mathsf{c_3d_4}\geq 0$) and, (ii) its output trace is less than or equal to the the identity ($\hat A_\mathsf{a_1b_2}^\mathsf{c_3d_4} \hat I_\mathsf{c_3d_4} \leq \hat I_\mathsf{a_1b_2}$).   Inputs and outputs define a direction in time. Space, in the circuit framework, is written into the fact that we can have composite systems. The first condition is consistent with time symmetry since positivity of the input transpose is equivalent to positivity of the output transpose (e.g.\ $\hat A_\mathsf{a^T_1b^T_2}^\mathsf{c_3d_4}\geq 0$ is equivalent to $\hat A_\mathsf{a_1b_2}^\mathsf{c^T_3d^T_4}\geq 0$). However, this condition does put space and time on a slightly different footing.  This is analogous to the way that time is treated in a slightly different way in relativity (the time coordinate is associated with, say, a minus sign in the metric while the space coordinates are associated with a plus sign).    Condition (ii) is equivalent to the condition that $\hat A^\mathsf{a_1g_3} \hat B_\mathsf{a_1}^\mathsf{b_2} \hat I_\mathsf{b_2g_3}\leq 1$ for any rank one projector $\hat A^\mathsf{a_1g_3}$.  The time reversed condition (which we do not demand) would read $\hat I^\mathsf{a_1g_3} \hat B_\mathsf{a_1}^\mathsf{b_2} \hat C_\mathsf{b_2g_3}\leq 1$ for any rank one projector $\hat C_\mathsf{b_2g_3}$.  It is easy to see that $\hat B_\mathsf{a_1}=\hat I_\mathsf{a_1}$ is physical.  On the other hand, $\hat B^\mathsf{b_2}=\hat I^\mathsf{b_2}$ is not physical although it does satisfy the time reversed condition just given.  Axiom 2 implies that there exists an operation corresponding to any physical operator.  Hence, there is an operation corresponding to $\hat I_\mathsf{a_1}$.  There cannot be an operation corresponding to $\hat I^\mathsf{b_2}$, however, as it is nonphysical.  Thus we see that this formalism is explicitly time asymmetric and that this asymmetry comes from condition (ii) in the definition of physicallity.

In the case that we restrict ourselves only to pure states and unitary transformations this time asymmetry disappears.  However, unitary transformations are not sufficient to cover the cases where we extract non-trivial measurement results (i.e.\ they fail to cover any non-trivial experiment).  It is an important question as to whether this formalism implies a fundamental time asymmetry in nature or whether this is only apparent.  In this connection it is worth noting that the recent formulation of Leifer and Spekkens \cite{leifer2011formulating}, which shares some features with the present formalism, does not appear to be time asymmetric.

\subsection{Quantum field theory and quantum gravity}

The formalism here pertains to discrete circuits.  If we take a region of space time and cut it up into some number of pieces then we can place wires between these pieces to form a circuit.  Now, if these smaller pieces have only space-like (or null) boundaries then we can think of the future facing boundaries as corresponding to outputs and the past facing boundaries as corresponding to inputs. If some part of the boundary is time-like then it will correspond to both input and outputs (this is like a fragment having inputs and outputs coming out the side).  The challenge of setting up quantum field theory is to work out how to take the limit of this situation to the infinitesimal (rather than discrete) case.  However, this framework offers certain advantages as an approach to quantum field theory. Namely, it provides a formulation which is in keeping with the spirit of special relativity without necessary reference to any specific foliation.

This framework might also provide a good stepping stone to a theory of quantum gravity.  Formalism locality, as a desirable property, was motivated by considerations from quantum gravity \cite{hardy2010bformalism}. In standard quantum theory we use \lq\lq rectangular" regions of spacetime resulting from evolving a state from a state on a past space-like hypersurface to a future space-like hypersurface.  However, in a theory of quantum gravity we expect to have indefinite causal structure so it will be impossible to refer our laws to specially shaped regions of space time (the notion of \lq\lq space-like" will not make sense if we have indefinite causal structure).   If this reasoning is correct, formalism locality will be a necessary feature of quantum gravity. The limiting case of a theory of quantum gravity in the direction of applicability of quantum theory is likely to be a formulation of quantum theory with the formalism locality property.

One great challenge facing applying these techniques to quantum field theory and, possibly, to quantum gravity, is to know how to adapt or reproduce that relevant physics which is usually formulated in terms of differential equations using the Turing inspired ideas of computer science.





\section*{Acknowledgements}
\addcontentsline{toc}{section}{Acknowledgements}

I am very grateful to Chris Fuchs, Giulio Chiribella, Mauro D'Ariano, Paulo Perinotti, Tony Short, Rob Spekkens, Matthew Leifer, Markus Mueller, Bob Coecke, Samson Abramsky,  Bill Edwards, Prakash Panangaden, and Vanessa Hardy for various remarks that influenced this paper.  I am also grateful to the staff at the Casa Mia Cafe in Waterloo for providing a hospitable environment where much of this work was done.

Research at Perimeter Institute for Theoretical Physics is supported in part by the Government of Canada through NSERC and by the Province of Ontario
through MRI.

\bibliography{quantum}
\bibliographystyle{plain}

\end{document}